\documentclass[sigconf, nonacm]{acmart}

\newcommand\vldbdoi{XX.XX/XXX.XX}

\newcommand\vldbavailabilityurl{}
\newcommand\vldbpagestyle{plain}

\usepackage{caption}
\usepackage{multirow}
\usepackage{colortbl}
\usepackage{hhline}
\usepackage{subfig}
\usepackage{longtable}
\usepackage{graphicx}
\usepackage{balance}

\begin{document}
\title{An Adaptive Column Compression Family for Self-Driving Databases}

\author{Marcell Feh\'{e}r}
\affiliation{%
  \institution{Agile Cloud Lab, Department of Electrical and Computer Engineering, DIGIT, Aarhus University}
  \city{Aarhus}
  \country{Denmark}
}
\email{sw0rdf1sh@ece.au.dk}

\author{Daniel E. Lucani}
\affiliation{%
  \institution{Agile Cloud Lab, Department of Electrical and Computer Engineering, DIGIT, Aarhus University}
  \city{Aarhus}
  \country{Denmark}
}
\email{daniel.lucani@ece.au.dk}

\author{Ioannis Chatzigeorgiou}
\affiliation{%
  \institution{School of Computing and Communications,\\Lancaster University}
  \city{Lancaster}
  \country{United Kingdom}
}
\email{i.chatzigeorgiou@lancaster.ac.uk}

\begin{abstract}
Modern in-memory databases are typically used for high-performance workloads, therefore they have to be optimized for small memory footprint and high query speed at the same time. Data compression has the potential to reduce memory requirements but often reduces query speed too. In this paper we propose a novel, adaptive compressor that offers a new trade-off point of these dimensions, achieving better compression than LZ4 while reaching query speeds close to the fastest existing segment encoders. We evaluate our compressor both with synthetic data in isolation and on the TPC-H and Join Order Benchmarks, integrated into a modern relational column store, Hyrise. 
\end{abstract}

\maketitle

\pagestyle{\vldbpagestyle}

%\begingroup\small\noindent\raggedright\textbf{PVLDB Reference Format:}\\
%\vldbauthors. \vldbtitle. PVLDB, \vldbvolume(\vldbissue): \vldbpages, \vldbyear.\\
%\href{https://doi.org/\vldbdoi}{doi:\vldbdoi}
%\endgroup
\begingroup
\renewcommand\thefootnote{}\footnote{\noindent
This work is licensed under the Creative Commons BY-NC-ND 4.0 International License. Visit https://creativecommons.org/licenses/by-nc-nd/4.0/ to view a copy of this license. For any use beyond those covered by this license, obtain permission by emailing info@vldb.org. Copyright is held by the owner/author(s). To Appear in the 13th Workshop on Accelerating Analytics and Data Management (ADMS'22), September 2022, Sydney, Australia.
}\addtocounter{footnote}{-1}
\endgroup

%\ifdefempty{\vldbavailabilityurl}{}{
%\vspace{.3cm}
%\begingroup\small\noindent\raggedright\textbf{PVLDB Artifact Availability:}\\
%The source code, data, and/or other artifacts have been made available at \url{\vldbavailabilityurl}.
%\endgroup

\section{Introduction}

% Compression is an important tool for in-memory databases
Database systems employ a wide variety of compression schemes to reduce memory footprint, increase effective storage capacity and overcome bandwidth limitations of slow hard drives. With the proliferation of in-memory databases, the role of data compression has changed from a nice-to-have feature to an essential tool that is required to store and analyze large datasets. Since the maximum amount of memory is an order of magnitude smaller than hard drive capacity in most desktop and server computers, simply adding more memory cannot solve the problem anymore.

% In cloud environments total price is driven by memory size of the instance
Additionally, more and more workloads move to the cloud where pricing models are dictated by the cloud provider. Hardware resources available for cloud-based virtual machines are typically pre-configured sets, where the price is proportional to the amount of memory and CPU included in each machine type. Therefore, efficient compression directly affects operational costs of in-memory databases running in cloud environments. 

% Column stores are more amenable to compression than row stores
Compression methods decrease data size by exploiting redundancy present within the input data, therefore column-oriented databases are more suitable to compression than traditional row stores. This is because a list of values from the same column always have the same data type and more likely to be compressible than records with multiple fields of different types. Modern in-memory column stores take this a step further and split each column to segments, where each segment has its own encoding and managed separately from the others. This approach enables great performance for both transactional and analytical workloads by using a write-optimized encoding of the most recent segment (where new data is added) and a read-optimized one for older segments. 
 
% Related Work (segment encoding selection)
A commonly used segment encoding strategy is assigning a single encoder per data type (integer, string, etc.) and use it for all segments in the database. While this decreases memory consumption and in some cases speeds up queries, it does not take into consideration the characteristics of the stored data. 
There is some research discussing dynamic encoding selection. CodecDB~\cite{codecdb} trains a model that infers which compressor is most likely to achieve the highest compression based on static metrics derived from the target column, such as cardinality or domain. Cen \textit{et al.} proposed a system called Learned Encoding Advisor~\cite{10.1145/3464509.3464885}, which takes into consideration the values (both statistics and a 1\% extract), sample queries and the underlying hardware as well, when selecting the best encoder. It is evaluated with two high-level strategy options: maximum compression and highest query performance. Boissier designed an encoding selection module for Hyrise, which learns cost models of encoders from the physical query cache, and provides the best segment encoder for a given memory budget~\cite{10.14778/3503585.3503588}.

% Overview of contributions of this paper (high level list)
In this paper we introduce a novel segment encoder family for integer columns which determines its own best parameters based on the data to be encoded. We evaluate them against the most commonly used segment compressors, both in isolation using synthetic data, and in a fully featured relational column store on the TPC-H~\cite{tpch} and Join Order Benchmarks~\cite{leis2015good}, analyzing the effects of different encoding parameters. Extending the idea of finding the best parameter for a single encoder to the whole database, we propose a segment encoding scheme for autonomous databases where the compressor selection is driven by query patterns to maximize performance.

% Paper organization
The paper is organized as follows. In \autoref{sec:background} we review the current state of segment encoding in modern relational databases and introduce the data compression technique that our encoders are based on. Then, in \autoref{sec:gd_segments} we present the detailed design of four segment variants using this compression algorithm. Our proposal for an intelligent, adaptive segment encoding framework for autonomous databases is presented in \autoref{sec:self_driving_db_segment_encoding}. The results of evaluating the new segment types against the commonly used ones, both in isolation and with industry-standard analytical benchmarks are shown in \autoref{sec:evaluation}. Finally, \autoref{sec:conclusions} draws conclusions from the results and proposes future work.

\section{Background}\label{sec:background}

\subsection{Segment Encoding in In-Memory Databases}

% related work on segment encoding
As companies collect and analyze increasingly large data sets, the performance of disk-based databases is not satisfactory anymore for many of them. With the decreasing price and increasing size and speed of main memory, in-memory databases have become viable and affordable alternatives~\cite{DBS-058} of disk-based systems. They are ideal for performance-critical workloads, which means processing speed and memory footprint are the new important metrics~\cite{manegold2000optimizing}. 
Compression has the potential to markedly decrease the footprint, especially in modern columnar databases like SAP HANA~\cite{saphana}, HyPer~\cite{hyper} or DuckDB~\cite{duckdb}, which horizontally partition data into segments and the unit of encoding is a segment instead of the whole column. Therefore, compressors enjoy two extra benefits in this environment: they can work on an array of values of the same type, and data distribution within a segment is more likely to be self-similar than in the whole column. Both of these qualities work in favor of data compressors. However, as decompression requires extra processing, it typically comes at a cost of query performance. Therefore, it is critical to select the most appropriate compressor for columns that are heavily used in queries. Since query performance in analytical workloads mostly depends on processing speed of integer columns~\cite{heinzl2021evaluating}, we focus on lossless integer compression. 
 
Note, that strings make up the majority of data in both real world datasets and widely used OLAP benchmarks like TPC-H and TPC-DS. In many cases people store numeric columns like timestamps, booleans or floats as strings as well, and cast them in the SQL query~\cite{vogelsgesangGetReal}. Therefore, integer compression usually has a minor effect on the overall storage footprint. However, there are several major use cases where numeric data is the dominant type, for example IoT time series. Consequently, our goal is to reduce data size and retain or increase speed of frequent operations on integer columns at the same time. 

We assess our proposed segment encoders against the most commonly used integer compression schemes in columnar databases.

\begin{itemize}
\item Dictionary Encoding~\cite{abadi2006integrating} is a widely used data compression technique where unique values are collected to a sorted dictionary, and their occurrences in the input data vector are represented by a list of dictionary offsets. It is the default encoding scheme of several column stores.
\item Frame-of-Reference (FoR)~\cite{goldstein1998compressing} encoding stores difference of each value to the common minimum, as well as the minimum itself. The delta values are bit packed to the smallest possible width. For evaluations we used an improved version is this technique, called Patched FoR (or PFoR). It splits the input data to fixed size blocks and performs FoR on each one separately, hoping to exploit local similarities present in the (large) array. 
\item LZ4~\cite{lz4} is a relatively heavy statistical compressor that offers high degree of data size reduction for slower compression and decompression than the other two lightweight methods. This scheme is typically used to compress columns that never or very rarely participate in query conditions.
\end{itemize}

% Hyrise
To test our segment compressor in a real system, we use Hyrise~\cite{dreseler2019hyrise}, a modern relational in-memory column store. It is an open-source research database with an extensible framework for segment encoders. All three encodings above are standard built-in options, making it easy for a segment developer to run benchmarks against them.
Even though vectorized implementations of some of these algorithms are available via libraries~\cite{10.1145/2735629}, we are using the non-SIMD variants for our comparisons.

\subsection{Generalized Deduplication}\label{sec:gd}

% Dictionary encoding 
Dictionary encoding, paired with null suppression of the dictionary, attribute vector or both, achieves high compression when the cardinality of the input data is sufficiently low (e.g., there are few unique values). However, when applied to datasets where all values are different, like a primary key column in a relational database, it inflates the data and slows down every operation. Several methods have been developed for compressing high cardinality datasets, each with their own expectations about the value distribution. For example, FoR and PFoR encodings assume a narrow domain of encoded values. 

\begin{figure}[ht]
\includegraphics[width=\columnwidth]{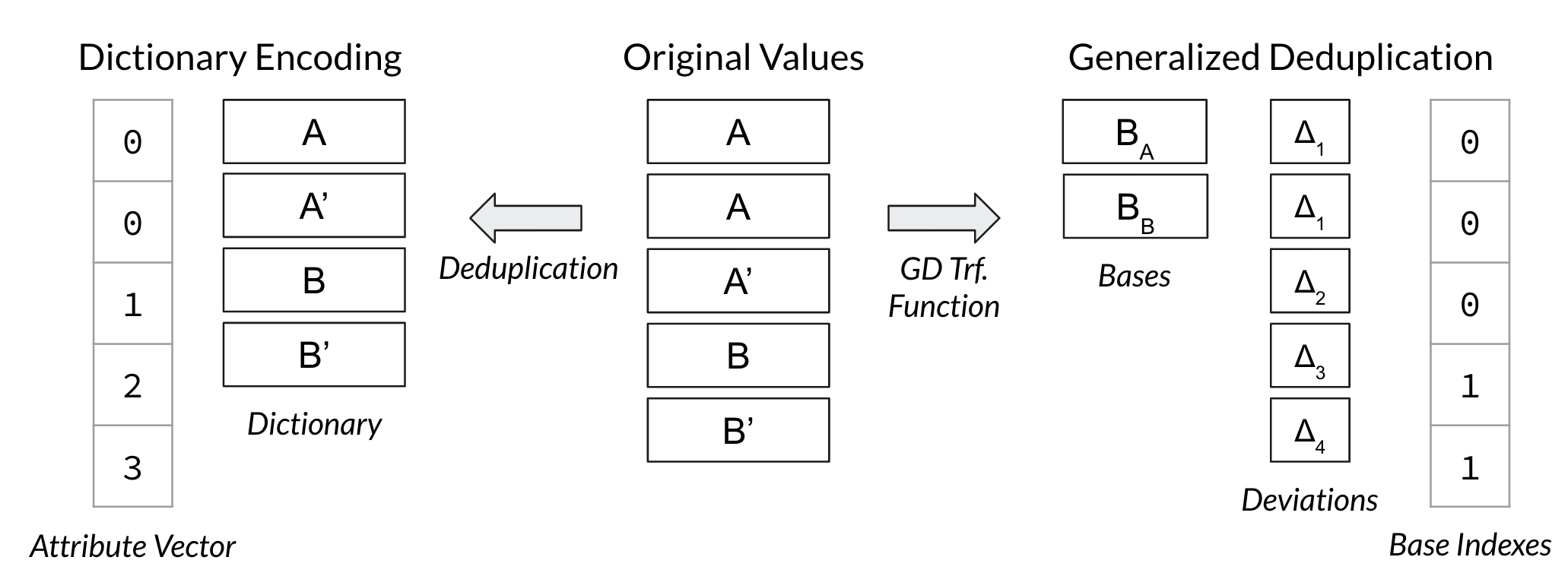}
\caption{Dictionary Encoding and Generalized Deduplication of the same set of values. GD uses an arbitrary, user-specified transformation function to convert input values to bases and deviations.}
\label{fig:gd_vs_dict}
\end{figure}

\begin{figure}[ht]
\includegraphics[width=\columnwidth]{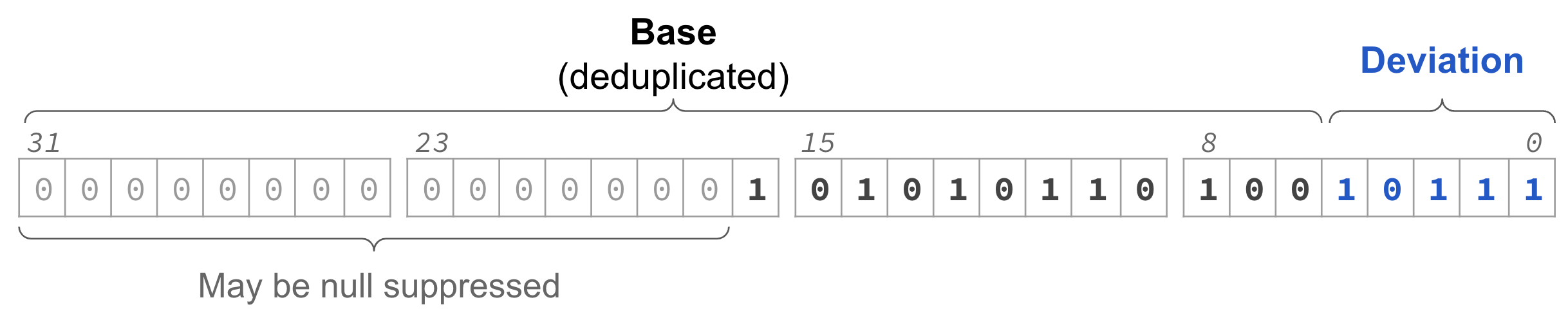}
\caption{Decomposition of a 4-byte unsigned integer (87703) to base (2740) and deviation (23) using the LastBit transformation with 5 bit deviations.}
\label{fig:gd_lastbit_number}
\end{figure}

% Generalized Deduplication
A new technique, introduced by Vestergaard \textit{et al.} ~\cite{vestergaard2019generalized} and referred to as \textit{Generalized Deduplication} (GD) does not have any inherent assumptions about the data distribution. It prodives a flexible compression framework that can be tailored for the data at hand. Compression with GD is a two-step process: first, every input value is passed through a user-defined, arbitrary \textit{transformation function} that converts it to a pair of \textit{base} and \textit{deviation}, where similar inputs should generate the same base with different deviations. The goal of this function is to separate the identical and varying parts of the input data chunks. Think of it as a booster step in dictionary encoding, which aims to increase the deduplication rate. Secondly, bases are deduplicated and the base index of each original value is determined (see \autoref{fig:gd_vs_dict}). Decompression is a straightforward process in the opposite direction: the base and deviation is looked up and passed to the inverse transformation function, which reconstructs the original value.
Generalized Deduplication has the following key properties:
\begin{itemize}
    \item It can be either a lossless or a lossy compressor, depending on whether the deviations are kept or discarded.
    \item Dictionary Encoding is a special case of GD, where the transformation  is the identity function.
    \item GD supports constant time random access, since reconstructing a single value given its index requires 3 lookups: the base index, the deviation and the base itself (using the base index).
    \item Arbitrary assumptions about the input data characteristics can be encoded as the transformation function. 
    \item Null suppression (removing leading zeroes, see \autoref{fig:gd_lastbit_number}) can be applied to any combination of the bases, deviations and base indexes.
    \item The customizability of GD makes it a good fit in a variety of use cases, including file systems, cloud storage, time series or relational databases. It is applicable in every scenario when the similarity between input data chunks can be extracted with a reversible transformation.
\end{itemize}

\begin{figure}[h]
  \includegraphics[width=\columnwidth]{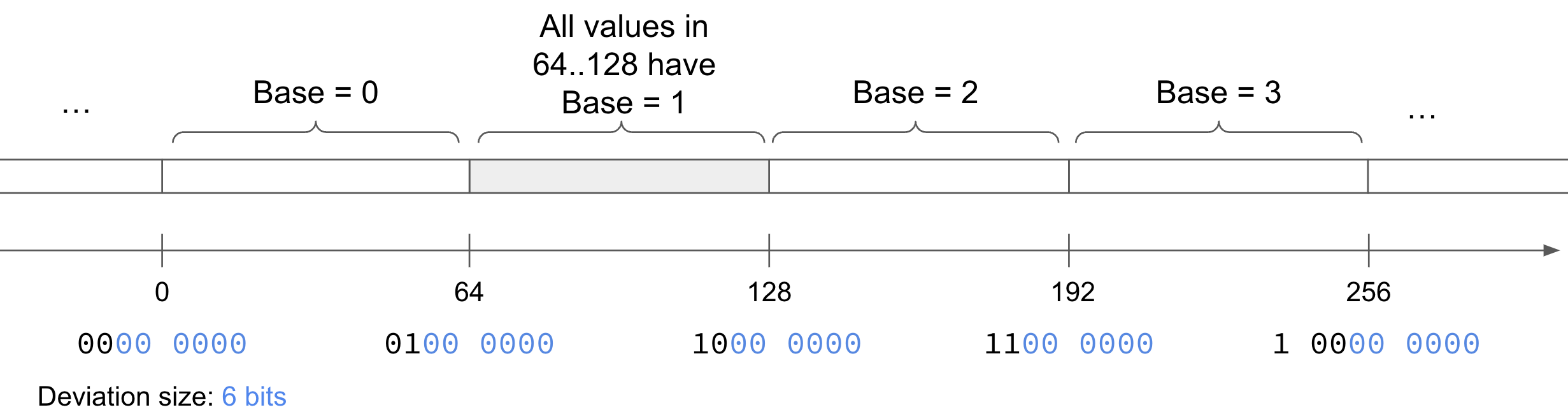}
  \caption{Partitioning integers with LastBit transformation using 6-bit deviations.}
  \label{fig:gd_lastbit_partitions}
  \end{figure}

% Transformation function
GD-based compressors have been proposed in the past for different data and use cases (e.g.,~\cite{feher2020smartmeter,9603654}), but not as a segment encoder in relational databases.
For integer compression in this environment, we chose a simple but powerful transformation function that considers the \textit{n} least significant bits of a data value as deviation, and the previous bits as base (see \autoref{fig:gd_lastbit_number}). When applied to a series of values, this transformation assumes they are close to each other, e.g. the bits with the highest variation across the whole segment are at the end of the values. This transformation function, called \textit{LastBit}, partitions the range of integers to consecutive regions of size $2^n$, that is, $[i2^n...(i+1)2^n-1]$ for $i=0,1,...$, where the base of every value within region $i$ is $i2^n$ (the smallest element of the region). For example with 6 bit deviations each base covers 64 consecutive values (see \autoref{fig:gd_lastbit_partitions}). It has an additional advantage besides simplicity (therefore, processing speed): it is order preserving for both bases and deviations, e.g., if $x > y$ then either $Base(x) > Base(y)$ or $Base(x) = Base(y)$ and $Deviation(x) > Deviation(y)$. This property gives LastBit an advantage over other transformation functions in relational databases, since it enables faster predicate evaluations in certain cases.

\begin{figure*}[t]
\includegraphics[width=\textwidth]{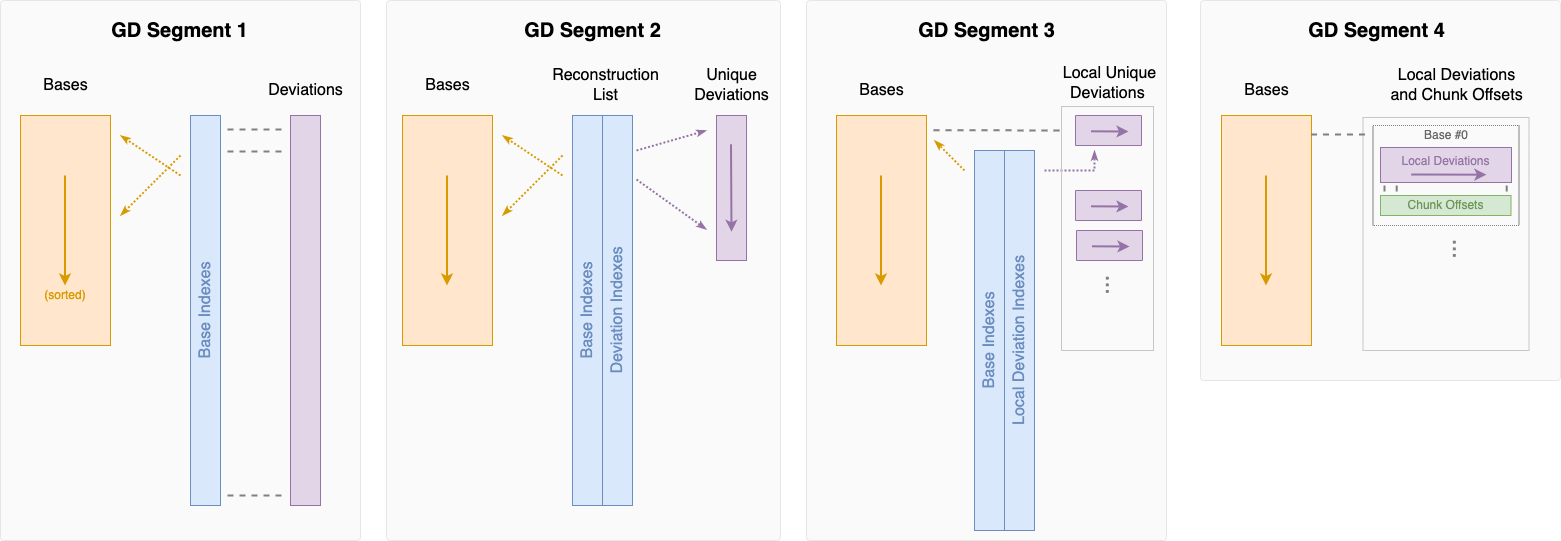}
\caption{Proposed GD-based segment encoders.}
\label{fig:gd_segments_design}
\end{figure*}

\section{GD Segments}\label{sec:gd_segments}

We propose four designs based on Generalized Deduplication with the LastBit transformation function to store, and access and query numeric segments in a column store (\autoref{fig:gd_segments_design}). When a segment is encoded, each value is split into a base and a deviation. The deviation size is an input parameter of the segment constructor, given in bits. 
All segment variants include the deduplicated and sorted list of bases, but they store deviations and base indexes in different ways. The segments are implemented in C++17, where we used \texttt{compact::vector}\footnote{\href{https://github.com/gmarcais/compact_vector}{https://github.com/gmarcais/compact\_vector}} for bit-packed lists and \texttt{std::vector} for regular ones. Bases and deviations are bit-packed lists in all four variants, using exactly [32-deviation size] bits per base and [deviation size] bits per deviation.

\textbf{GD Segment 1} has the simplest internal structure where each deviation is stored separately. Both deviations and base indexes contain one entry per segment value, in the same order as the original data. For example, to reconstruct the first value of the segment, we need to combine the base indicated by the first base index with the first deviation. Base indexes is a bit-packed list just like the bases and deviations, but its width is determined in run-time to fit the largest index.

In \textbf{GD Segment 2} the deviations are also deduplicated and stored in a sorted list. Since we cannot find the deviation of value $i$ at $deviations[i]$ any more, we must store the deviation index for each value offset explicitly. To minimize the number of memory lookups during a reconstruction, we store each base index - deviation index pair in a single value, and construct a bit-packed list of them. The width of both indexes (and therefore the width of the reconstruction list) is calculated in run-time, based on the number of unique bases and deviations present in the segment.

\textbf{GD Segment 3} aims to reduce the number of bits needed to address deviations. Since in GD Segment 2 we stored all deviations in a single list, the deviation index in the reconstruction list must be wide enough for the largest one. Instead of a single list with all (unique) deviations, we can organize them for each base separately. We simply scan all base-deviation pairs produced by the segment values, and group the deviations by their bases. Finally, we remove the duplicates and sort. The resulting 2-dimensional array consists of the unique deviations for each base, which are shorter than the single global list of deviations, requiring fewer bits to address. Given that each base is associated with a local group of deviations, some deviations are likely to be members of multiple local groups and be stored multiple times.

\textbf{GD Segment 4} has a drastically different structure that is geared towards fast table scans but is no longer random accessible. Similarly to GD Segment 3, deviations are collected per base and sorted, but in this case not deduplicated. For each deviation we also store the original value offset, referred to as chunk offset on \autoref{fig:gd_segments_design}, which is needed to produce the result for table scans. 

The design choices of the GD Segment variants are motivated by the characteristics of generalized deduplication, the LastBit transformation and our goal to optimize for compression, random access and table scan speed at the same time. The four presented designs stand at different trade-off points between these dimensions.

% How is it possible to speed up table scan with GD Segments
\subsection{Fast Table Scans}

Query performance of a segment compressor primarily depends on the speed of two operations: random access (reconstruct the original value given its offset, also known as \textit{dereferencing}) and table scan (given a predicate and a query value, return the segment offsets that satisfy the condition). In \autoref{table:gd_segments_comparison} we have provided the formulas and number of memory lookups per random access, which directly affects random access speeds. Scans can be performed without any special support from the segment encoder, simply by iterating or decompressing the whole segment first, and evaluating the predicate on each value, one by one. While this yields correct results and is a reasonable default behavior, it can be slower than exploiting the internal structure of the encoded representation, if possible. As we already hinted, all GD Segment versions offer custom table scan implementations that do not require decompressing raw values for predicate evaluation. In fact, not a single value needs to be reconstructed to serve scans with any predicate in either of the GD Segment variants.

\begin{figure}[ht]
\includegraphics[width=\columnwidth]{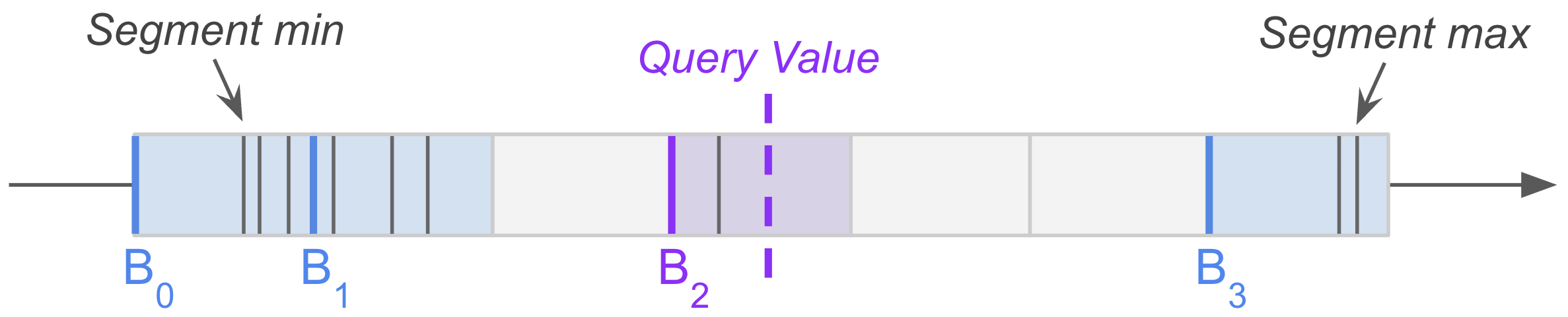}
\caption{A sample segment with only 9 stored values and 4 bases}
\Description{}
\label{fig:tablescan_illustration}
\end{figure}

The key to evaluating predicates\footnote{ The following predicates are valid for integer columns: Equals, NotEquals, Greater, GreaterEquals, Less, LessEquals} without decompressing a GD Segment is the LastBit transformation function. Since it partitions integers to disjunct regions, we can quickly determine which base ranges may contain values that satisfy the current query. \autoref{fig:tablescan_illustration} depicts a small segment containing only 9 values that map to 4 bases in total. When a table scan is performed, first the query value is passed through the same GD transformation function as the segment values did earlier, determining the query base and deviation. 
Next, searching for the query base in the bases list using \texttt{std::lower\_bound} (or \texttt{std::upper\_bound} depending on the predicate) tells us whether it is present in the segment, its index if present, or the index of the first larger base. Since the list of bases is sorted, we gain all this valuable information for a cheap, $O(log n)$ binary search. If the query base is present in the segment, this is the only base range that requires further investigation to evaluate the predicate. Values of all other base ranges are either completely contained in the result set, or not at all. For example, if the query predicate on \autoref{fig:tablescan_illustration} is \textit{GreaterEquals}, surely no value from base ranges $B_0$ and $B_1$ is a match, but every value of $B_3$ is. Whether a non-query value base range is included or excluded from the result set depends on the predicate. This zero-cost elimination can be done with all predicates, as it is a consequence of the order preserving nature of the LastBit transformation. 
If the query base is present in the segment, it needs to be scanned to find the values satisfying the predicate. Since deviations are also order preserving, there is no need to reconstruct the values of the query base. Comparing the stored deviations with the query deviation using the query predicate yields the correct results for the table scan. 

\begin{table*}[h!]
  \centering
  \begin{tabular}{c l c } 
   \textbf{Segment} & \textbf{Formula to reconstruct value at of offset $i$} \textit{(the GD inverse transformation is denoted by $\bigoplus$)} & \textbf{Lookups} \\
   \hline
   Uncompressed & \texttt{data[i]} & 1 \\
   Dictionary & \texttt{dictionary[attribute\_vector[i]]} & 2 \\
   GD Segment 1 & \texttt{bases[base\_indexes[i]]} $\bigoplus$ \texttt{deviations[i]} & 3 \\
   GD Segment 2 & \texttt{bases[base\_indexes[i]]} $\bigoplus$ \texttt{unique\_devs[deviation\_indexes[i]]} & 3 \\
   GD Segment 3 & \texttt{bases[base\_indexes[i]]} $\bigoplus$ \texttt{local\_unique\_devs[base\_indexes[i]] [local\_dev\_indexes[i]]} & 4 \\
   GD Segment 4 & \textit{not random accessible} & - \\
   \end{tabular}
  \caption{Random access formula and number of memory accesses for different segment types.}
  \label{table:gd_segments_comparison}
  \end{table*}

Determining which deviations belong to the query base and comparing them to the query deviation is where the four GD Segment designs differ the most. GD Segment 1 is the slowest, since it does not store the deviations per base, therefore the whole base indexes list must be traversed to find the deviation indexes that map to the query base. In GD Segment 2 the process is identical, but as only unique deviations are stored, there is a minor performance gain due to the better cache hit rate. GD Segment 3 is faster, because stored deviations have been grouped by bases, therefore it does not have to iterate through the whole reconstruction list when scanning the deviations of the query base. Additionally, since local unique deviations are sorted, finding which ones satisfy the query condition requires a fast binary search. GD Segment 4 is able to evaluate table scans with at most two binary searches due to its data layout, granting it exceptional speed. Unfortunately, without a reconstruction list, it also lost the ability of constant-time random access, since there is no way to look up the base and deviation based on the value offset. Instead, it has to iterate over all offset lists until the requested one is found.

%Note, that the total time to perform a table scan is dominated by memory allocation and writing the result set, rather than the efficiency of the algorithm, especially with high selectivity scans.

\section{Adaptive Segment Encoding}\label{sec:self_driving_db_segment_encoding}

Generalized deduplication is unique among the commonly used encoding schemes in the sense that there is no widely applicable default configuration that yields predictable performance independent of the data. For GD Segments, the deviation size that results in good compression and query performance heavily depends on the data distribution. A database administrator could pick an arbitrary size as a global default (e.g., 8-bit deviations), but it likely won't be the best setting across different columns, segments of the same column, or even the same segment over its whole lifetime. Thus, a fixed default almost certainly wastes memory and CPU cycles eventually. 

Instead of relying on an administrator to manually select the deviation size (either globally, or per-segment), we propose an iterative encoder for GD Segments. It automatically determines the best deviation size by trying all values from 1 to 30 bits (assuming 32-bit integer data) and selects the best one for the segment. The only problem is how to define the "best". It seems trivial to use the deviation size that achieves the highest compression, however, the best query performance is not necessarily at the same setting. This is due the different internal structures and custom table scan logic of GD Segment variants. The best option can only be determined based on testing the performance of different deviation sizes and knowing the relative importance of multiple factors: \textit{compression}, as well as the speed of \textit{random access}, \textit{sequential access} and \textit{table scan}. Even compression and decompression speeds are relevant in databases where segment encoding (and re-encoding) is a blocking operation. 

\begin{table}[b]
\centering
\begin{tabular}{l c c c c} 
 \textbf{Dev.Size} & \textbf{Comp.} & \textbf{Seq.Access} & \textbf{Rand.Access} & \textbf{TableScan} \\
 \hline
 1 bit & 2\% & 8 ns & 17 ns & 171 $\mu$s \\
 2 bits & 27\% & 8 ns & 16 ns & 160 $\mu$s \\
 3 bits & 39\% & 10 ns & 121 ns & 159 $\mu$s \\
 \multicolumn{5}{c}{...} \\
 30 bits & 3\% & 9 ns & 16 ns & 259 $\mu$s
 \end{tabular}
\caption{Sample results of GD Segment 1 diagnostic tests on a primary key segment}
\label{table:gd_segment_diag_results}
\end{table}

Our iterative GD Segment encoder works by first encoding the input data with all 30 possible deviation sizes. It records the compression and runs a series of speed tests to determine the performance of each option. We measure sequential access (by dereferencing all offsets), random access (by requesting a set of random offsets), and table scans using the six integer predicates and random query values. The result of diagnostic tests is a table similar to \autoref{table:gd_segment_diag_results}.

\begin{figure}[b]
\includegraphics[width=\columnwidth]{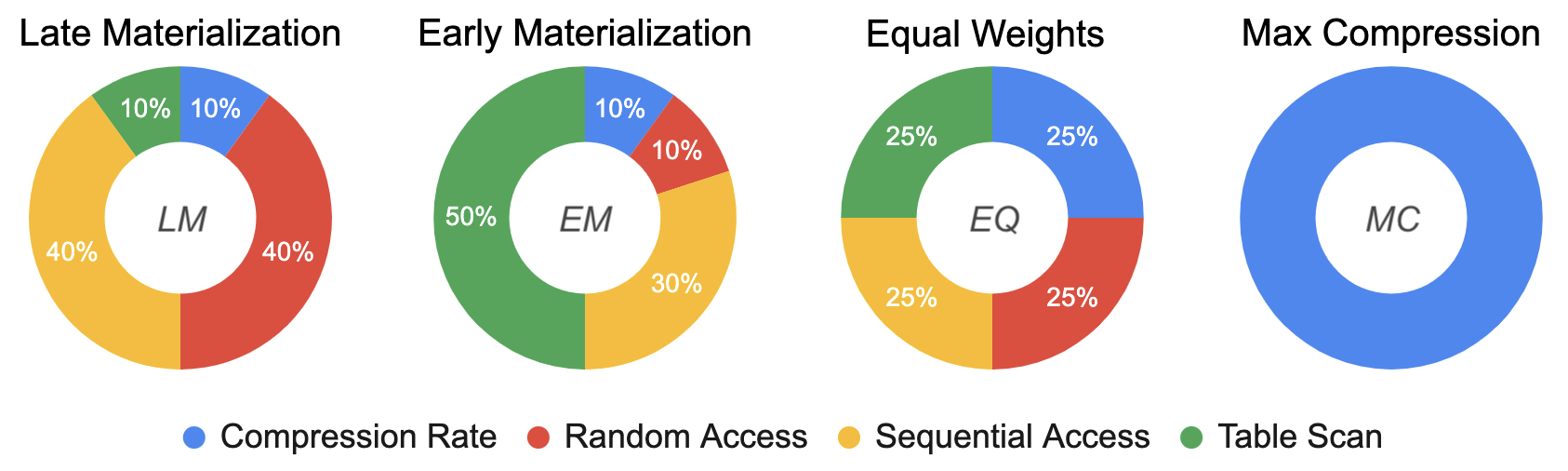}
\caption{Possible presets of relative importance between compression and query performance, when selecting the best segment encoder.}
\label{fig:encoding_weights}
\end{figure}

\begin{figure*}[t]
\includegraphics[width=\textwidth]{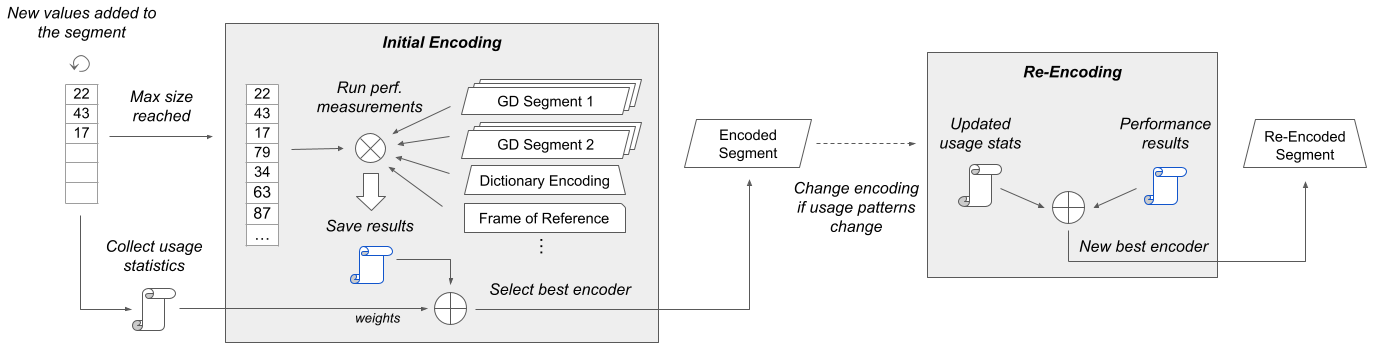}
\caption{Proposed adaptive segment encoding process for autonomous databases.}
\label{fig:segment_encoding_process}
\end{figure*}

Based on these measurements, there are multiple ways to select the best deviation size. If the relative importance of the dimensions is available, we can consider them as weights and find the deviation size that yields the best combination of compression and speeds. \autoref{fig:encoding_weights} shows a few examples of weight distributions. These can be fixed based on the database architecture, or even better, inferred from the workload experienced by the segment while it was unencoded.

Databases that use \textit{late materialization} (like Hyrise) execute queries in a way that lists of segment offsets are passed between processing nodes (e.g., a table scan or a join) and actual values are materialized as late as possible. As a result, random and sequential access via iterator dereferencing are the dominant segment operations. An alternative execution strategy is to materialize values at the very first operation (typically a table scan) and and pass them directly during evaluation. Therefore, \textit{early materialization} databases achieve better query performance if they select an encoding with more weight on table scan speed and less on random access.

Relational databases have long been collecting statistics to guide query planning. We propose extending this capability with high granularity access and table scan statistics on a per-segment basis. Specifically, recording the number of sequential and random accesses, as well as table scans with each predicate separately. If this detailed query history is available, the system can use it in the beginning of the segment encoding process to infer the relative importance of performance metrics, which provides the weights for selecting the most appropriate compressor configuration.

Moreover, performance-based encoding selection can be extended to all segment encoders and the complete lifetime of segments. A \textit{self-driving database}~\cite{pavlo2017self, ma2018query, ma2021self} is a relatively new concept which describes a system that automates maintenance and optimization tasks. If a database has up-to-date information about the usage patterns of each segment, it can assess whether the current encoding is still the best for the experienced workload. By running the performance measurements we proposed for finding the best GD Segment configuration, the database could compare every available encoder for every segment. In systems where compressed segment contents are immutable, the performance metrics stay valid throughout their whole lifetime. If usage patterns change over time, for example when data loses relevance and regular table scans are replaced by aggregate computations, the database can re-use the stored performance measurements together with the latest usage statistics in order to determine the best encoding going forward. It can also estimate the gains the new best encoding would yield over the current one, and decide whether it is worth re-encoding. This iterative process is illustrated by \autoref{fig:segment_encoding_process}. 

The evaluation of encoders can be performed at regular time intervals, or triggered when a new segment of a column is completed. When the system concludes that a segment is worth re-encoding, it can proceed immediately or schedule the job for a later time when the user-facing workload is expected to be low. Even though segment encoding is assumed to be a background operation, measuring different configurations for GD Segment (and presumably other encoders too) is computationally expensive and might indirectly affect system performance. Therefore, this initial profiling could also be deferred to a quiet period, if possible.

It is worth noting, that the best balance between compression and access/scan speeds cannot be derived automatically from the proposed performance metrics and usage statistics, but it is rather an arbitrary choice. The database user, administrator or system itself must decide how much speed they are willing to trade for a smaller memory footprint and, hence, a lower cost. Sometimes the same segment encoder configuration yields the best compression and highest speed at the same time. In this case the decision is trivial, but many times it is less so. For example, when a segment is used extremely rarely, search and access performance is much less important than size reduction, thus a heavy compressor like LZ4 is a better option than dictionary encoding, even though it has orders of magnitude worse operational scores. The same choice may be present between different configurations of the same segment encoder.

\section{Evaluations}\label{sec:evaluation}

We evaluated GD Segment against uncompressed, dictionary encoded, frame-of-reference encoded and LZ4 compressed segments in two scenarios. First, we measured their compression and query performance in isolation using synthetic datasets. Then we integrated GD Segment 1 into Hyrise and measured the TPC-H and Join Order Benchmarks. The purpose of these tests is twofold. First, to see how GD Segment variants compare to commonly used segment encoders on typical integer columns. We also wanted to see how important it is to select the deviation size based on the data and relative importance of different factors, e.g., what the performance gain is when the deviation size is chosen based on measurements versus constant 8 bits.

\subsection{Standalone Evaluation}

We have implemented dictionary, Frame-of-Reference and LZ4\footnote{Using the ZLIB library, which is natively available in every operating system} segment encoders in C++17 for standalone testing. They are functionally equivalent to their Hyrise counterparts: dictionary encoder uses a byte-aligned attribute vector and the same custom table scan logic, PFoR encodes the segment in blocks of 2048 values. The only difference in our implementation is that LZ4 segment compresses the whole segment as a single block (while in Hyrise it partitions the data to 16kB blocks), therefore a full segment decompression is required in the beginning of random access tests. Both LZ4 and PFoR segments fully decompress the values during table scans. GD Segment variants measure compression and query performance with all deviation sizes between 1 and 30 bits. We used the default segment size of Hyrise, which is 65535 elements. All standalone measurements were performed on a single thread of a 2.6 GHz Intel Core i7 CPU with 256 kB L2 and 12 MB L3 cache. The segments are stored and profiled entirely in memory.

\begin{figure*}[t!]
  \includegraphics[height=3.2cm,trim={0 1cm 31cm 0},clip]{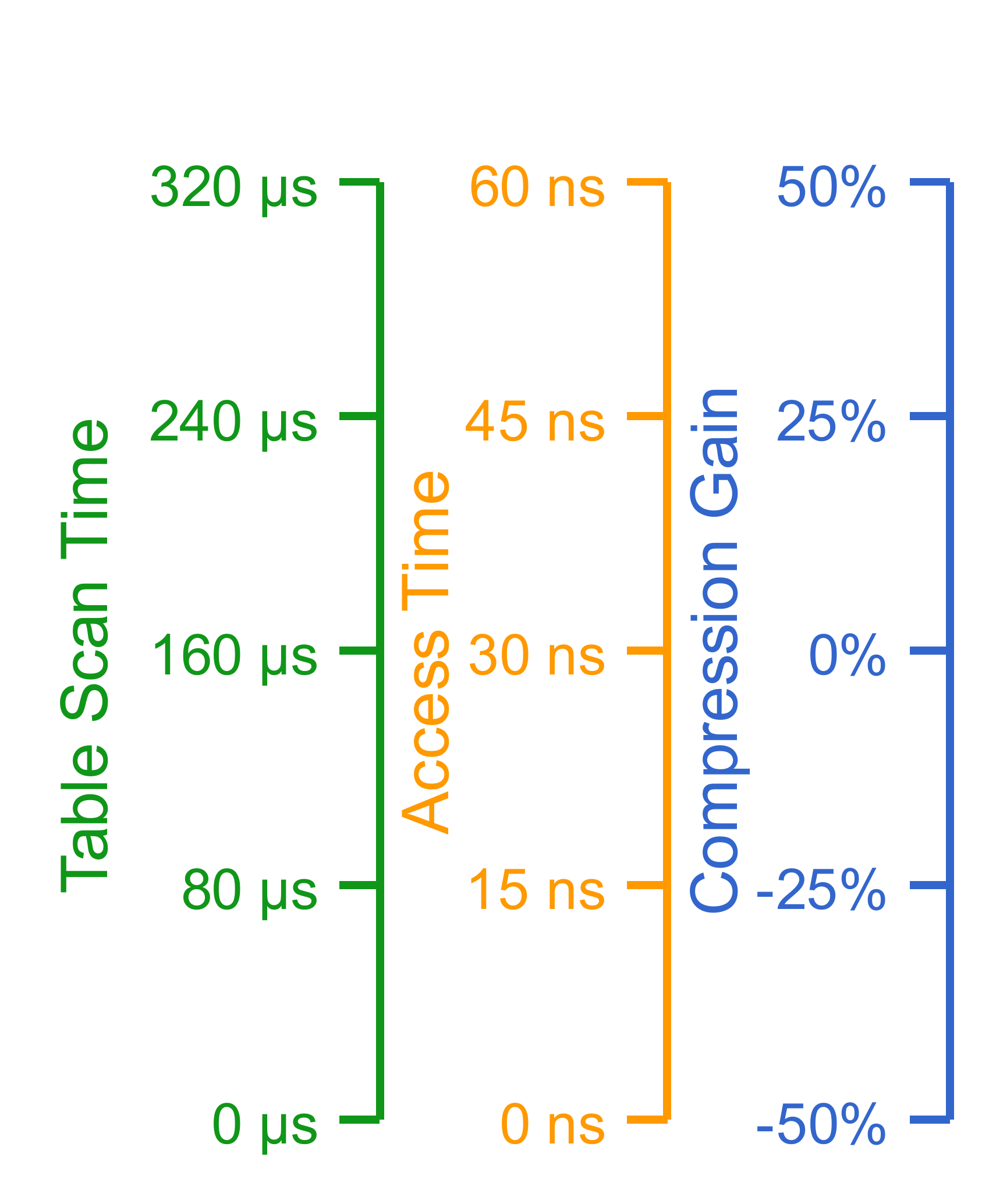}
  \includegraphics[height=3.2cm,trim={0 1cm 0 0},clip]{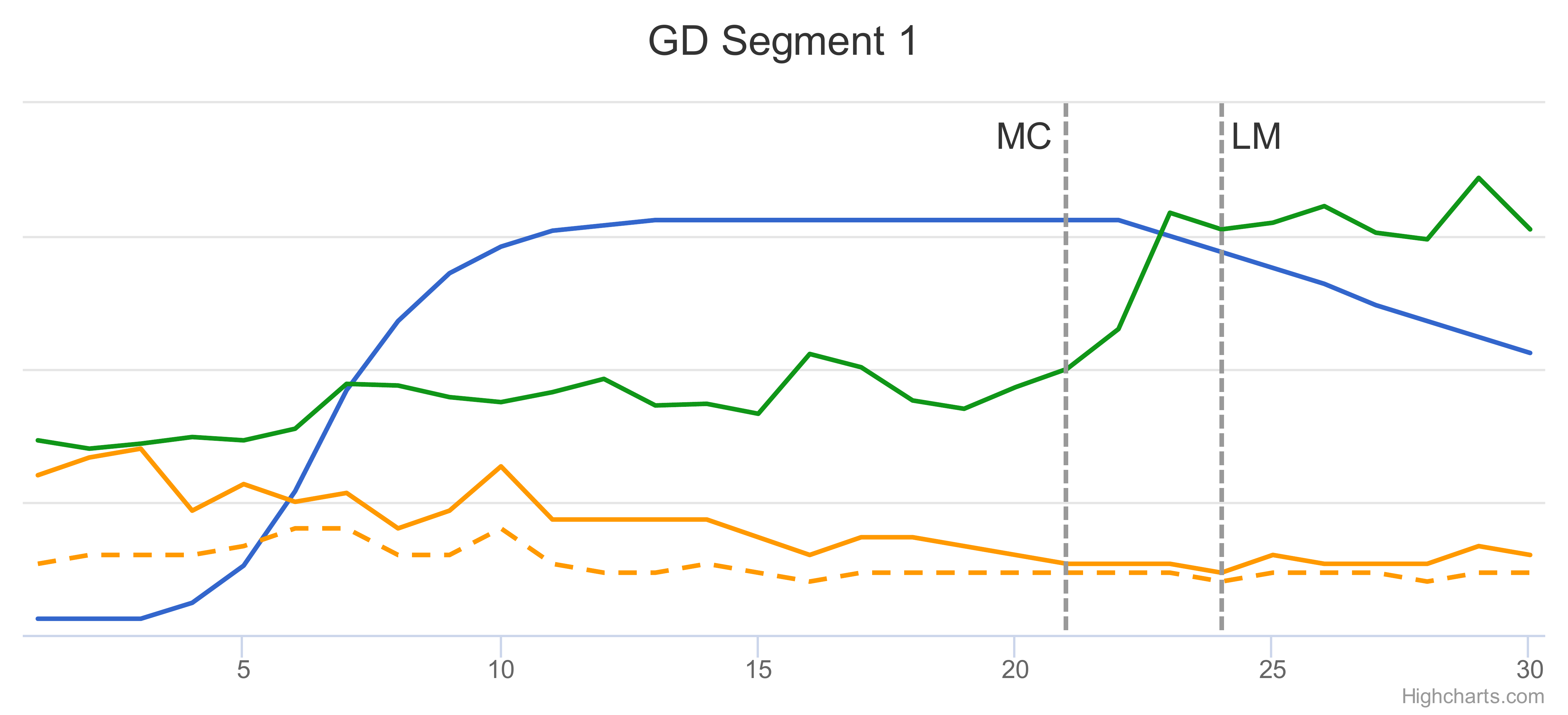}
  \includegraphics[height=3.2cm,trim={0 1cm 0 0},clip]{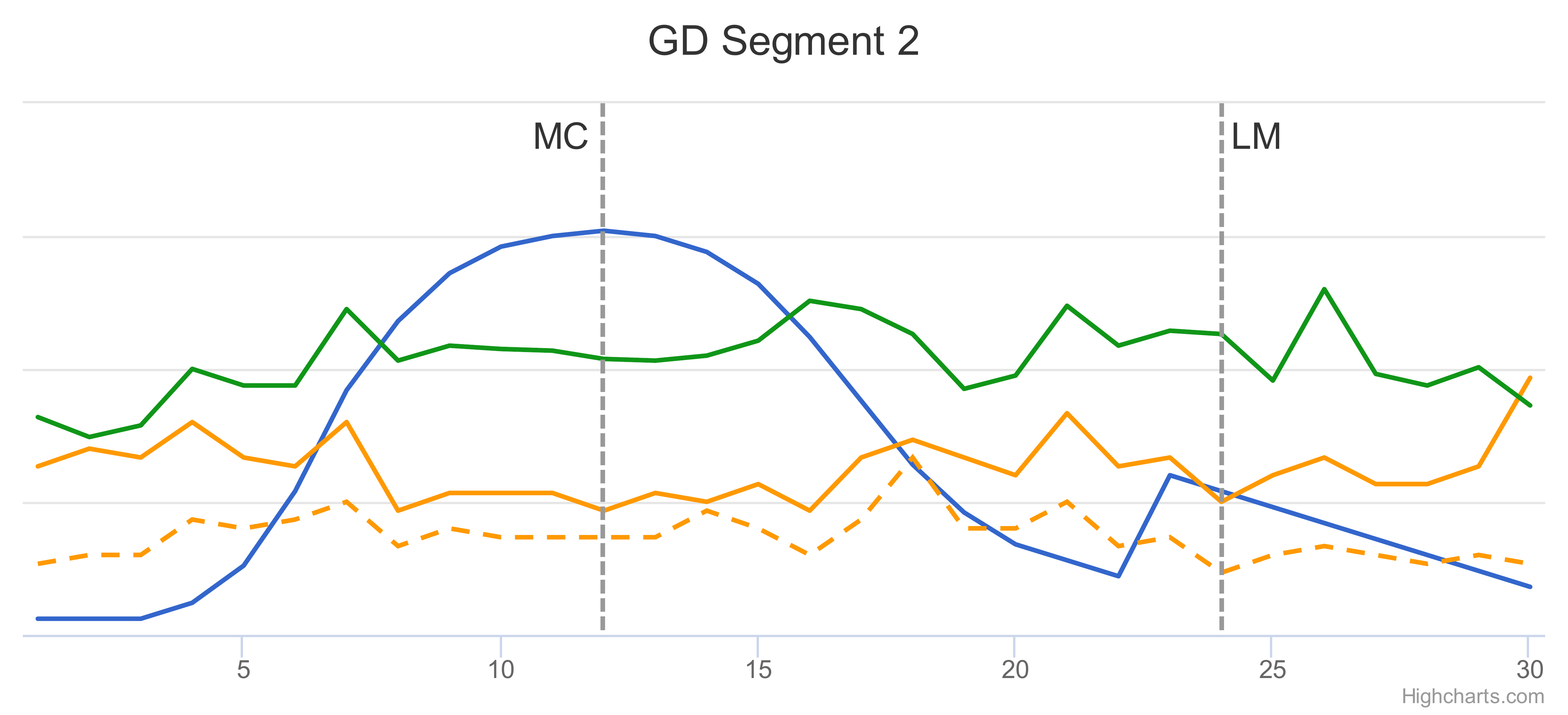}
\\
  \includegraphics[height=3.2cm,trim={0 1cm 30cm 0},clip]{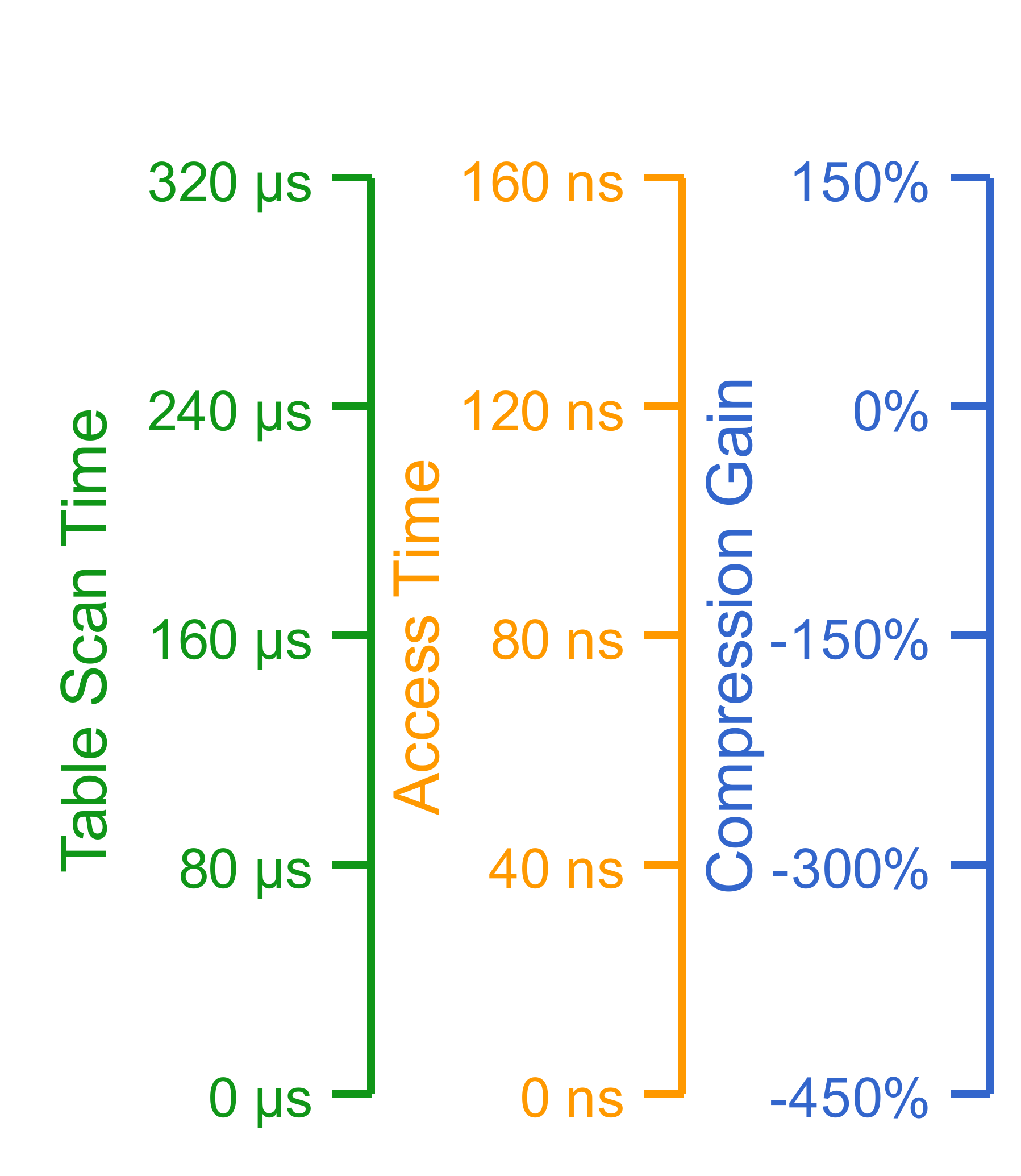}
  \includegraphics[height=3.2cm,trim={0 1cm 0 0},clip]{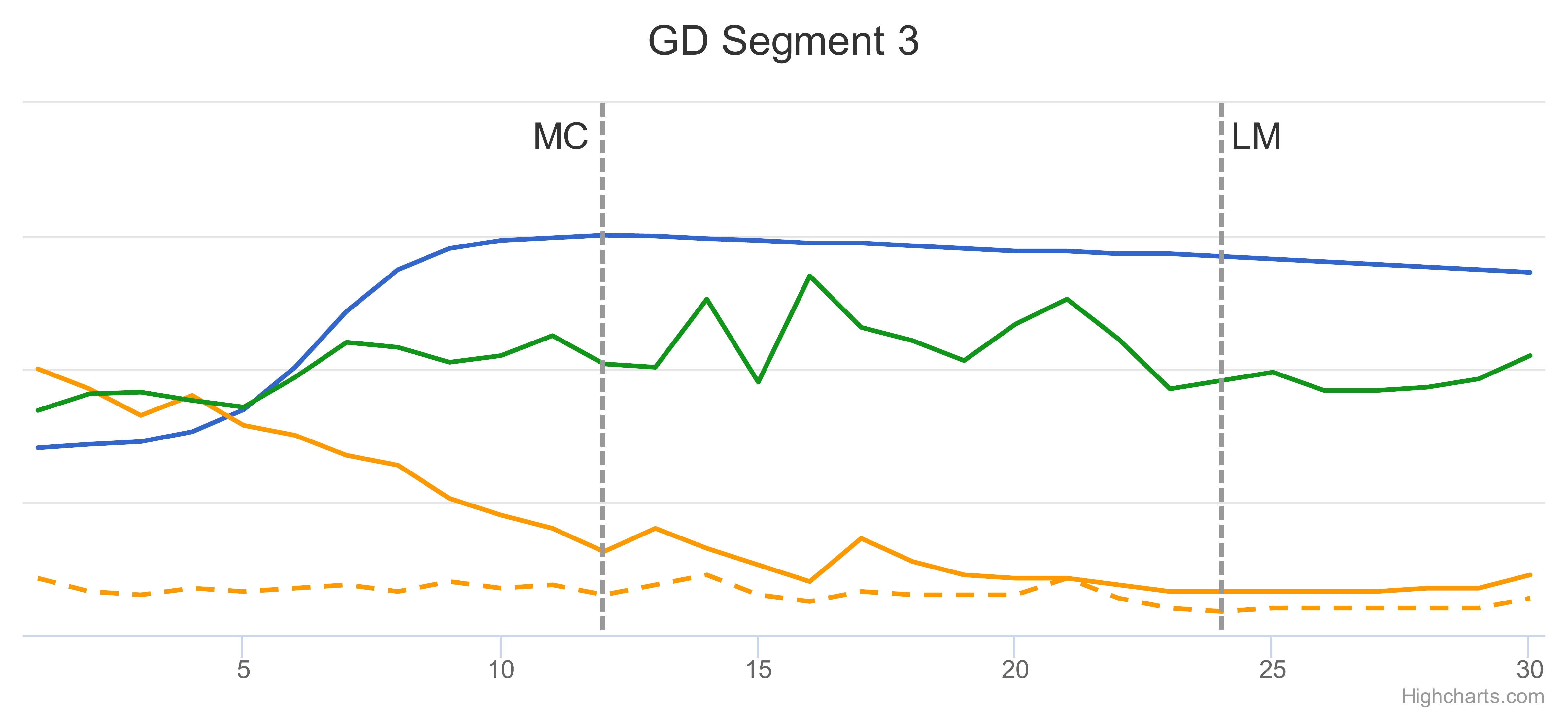}
  \includegraphics[height=3.2cm,trim={0 1cm 0 0},clip]{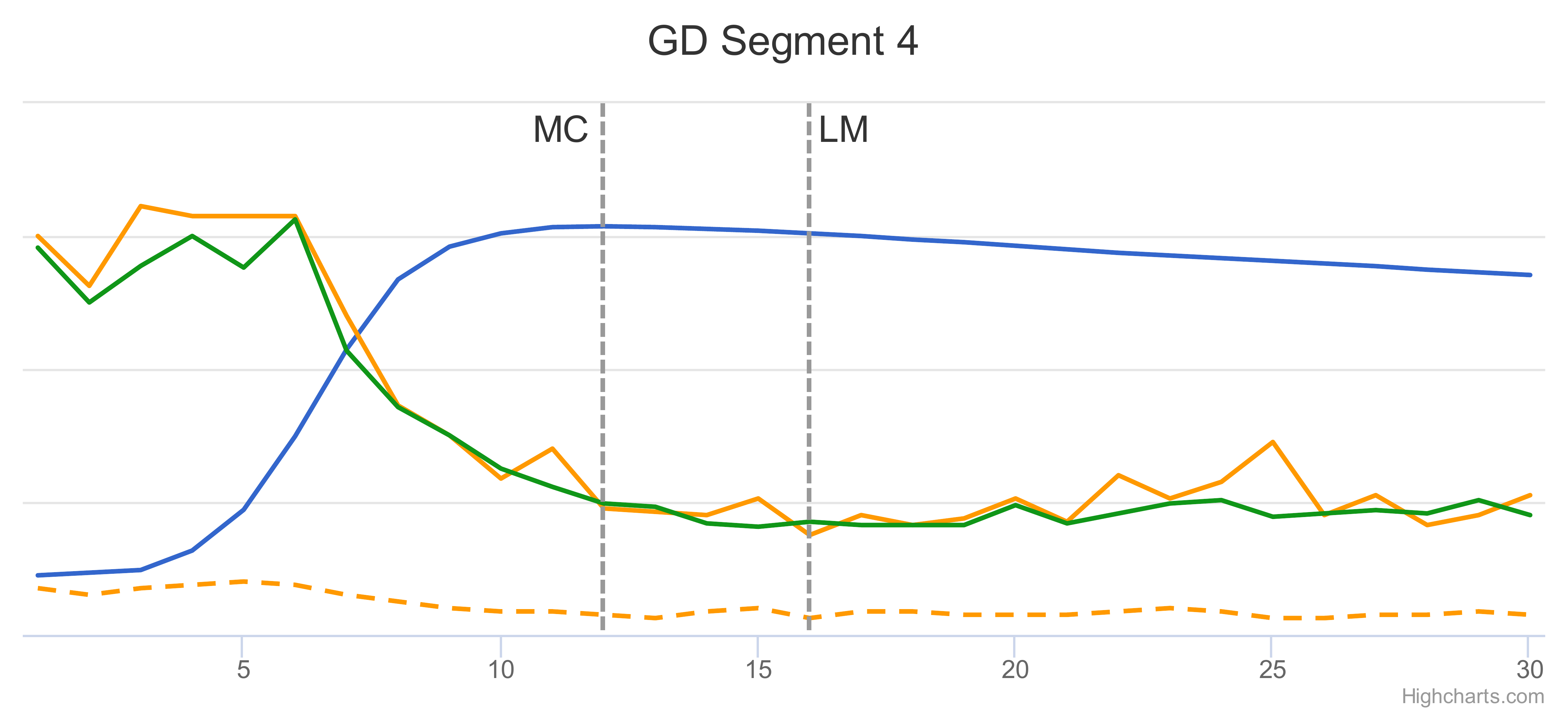}
  \caption{Compression gain (higher is better), access (dashed line: sequential, solid: random) and table scan times (lower is better) of the time series segment encoded with every GD Segment variant at the whole range of deviation sizes, between 1-30 bits ($x$ axis). The best deviation size of Maximum Compression (MC) and Late Materialization (LM) performance presets are marked. Note, that the access time and compression scales are different for GD Segments 1-2 and 3-4.}
  \label{fig:best_dev_size_time_series}
\end{figure*} 

We used the same synthetic datasets as Heinzl \textit{et al.} in~\cite{heinzl2021evaluating} with the addition of a primary key segment. The datasets are: (i) uniformly distributed random numbers between 0 and $2^{32}$, (ii) sorted equidistant numbers with the step of 5, (iii) years between 1900 and 2100, (iv) months between 1 and 12, (v) a time series (power consumption readings of a household) starting at $10^6$ and (vi) primary key starting with 1. 

Compression gain is reported as a percentage of decrease in size, e.g., 0\% indicates no compression and 50\% means the compressed segment is half the size of the original data vector. To measure random access time, we dereferenced a uniform random set of 6553 offsets (10\% of the segment size), and report the average time. For sequential access, all offsets from 0 to 65535 are requested and the average time is reported. Segments that do not support random access (LZ4 and GD Segment 4) decompress the whole segment and return elements of the reconstructed integer vector. Decompression time is included in the reported times for these two segment types. For table scans we generate a uniform random set of 655 query values (1\% of the segment size) in the range of the stored values with an extension of $\pm$ 10\% to simulate out-of-range scans. Then, we perform a table scan with all six predicates for every query value and report the average time of all 3930 scans. 

Note, that it would be straightforward to differentiate table scan times per predicate as separate performance metrics and allow the database to assign different weights to each one when determining the best encoding for a segment. We chose to aggregate all predicates only to ease the visualization of our results by decreasing the number of dimensions. We are also not reporting compression and decompression times, because they are irrelevant in a database where both initial encoding and re-encoding are non-blocking background operations. Therefore, encoding speed does not directly affect the user-perceived query performance. In use-cases where segment encoding is done synchronously, these should be part of the segment performance report as well.

%\begin{figure}[t]
%\includegraphics[width=\columnwidth]{figures/time_series_gd_variants_qualities.png}
%\caption{Compression gain (higher is better), access and table scan times (lower is better) of the time series segment encoded with each GD Segment variant at every deviation size between 1-30 bits ($x$ axis).}
%\label{fig:best_dev_size_time_series}
%\end{figure}

\autoref{fig:best_dev_size_time_series} illustrates how each metric changes when different deviation sizes are used to encode the time series segment. As a general trend it can be stated that when compression gain increases, both access and table scan times decrease. However, the deviation size that maximizes compression (marked MC) and expected performance of late materialization databases (marked LM) are different in every segment variant for this column. Therefore, we cannot assume that maximum compression automatically results in the best query performance.

\begin{figure}[ht]
  
  \includegraphics[width=\columnwidth,trim={0 1cm 0 0},clip]{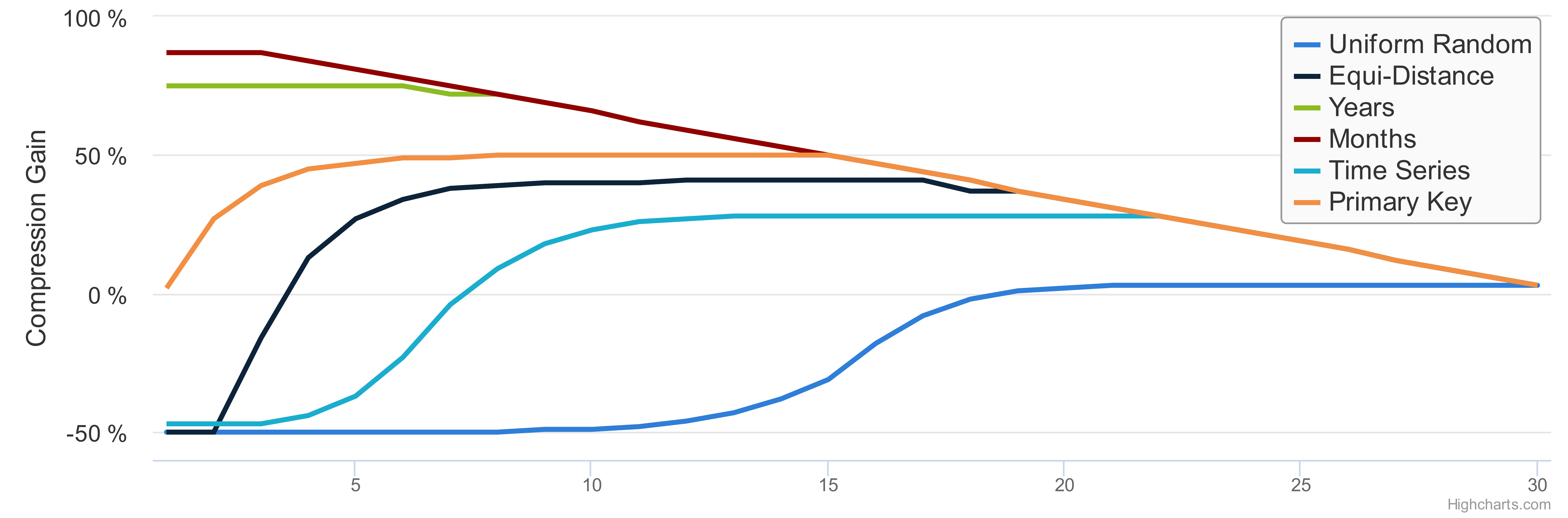}
  
  \vspace{.5cm}

  \includegraphics[width=\columnwidth,trim={0 1cm 0 0},clip]{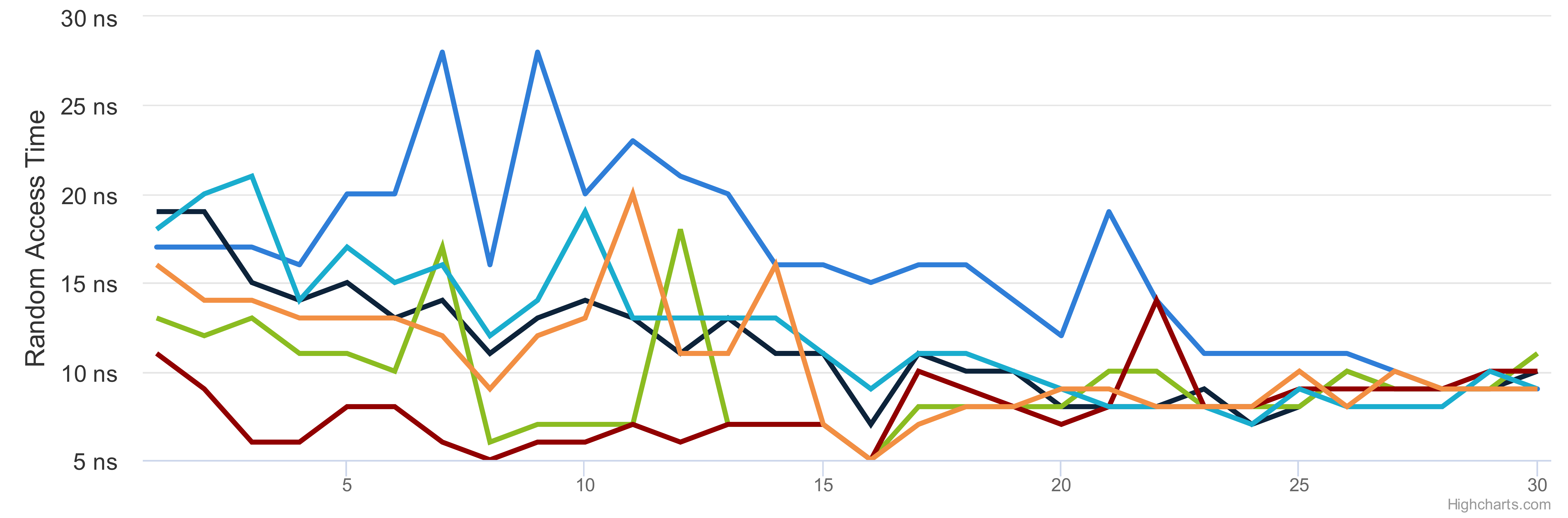}
  
  \vspace{.5cm}

  \includegraphics[width=\columnwidth,trim={0 1cm 0 0},clip]{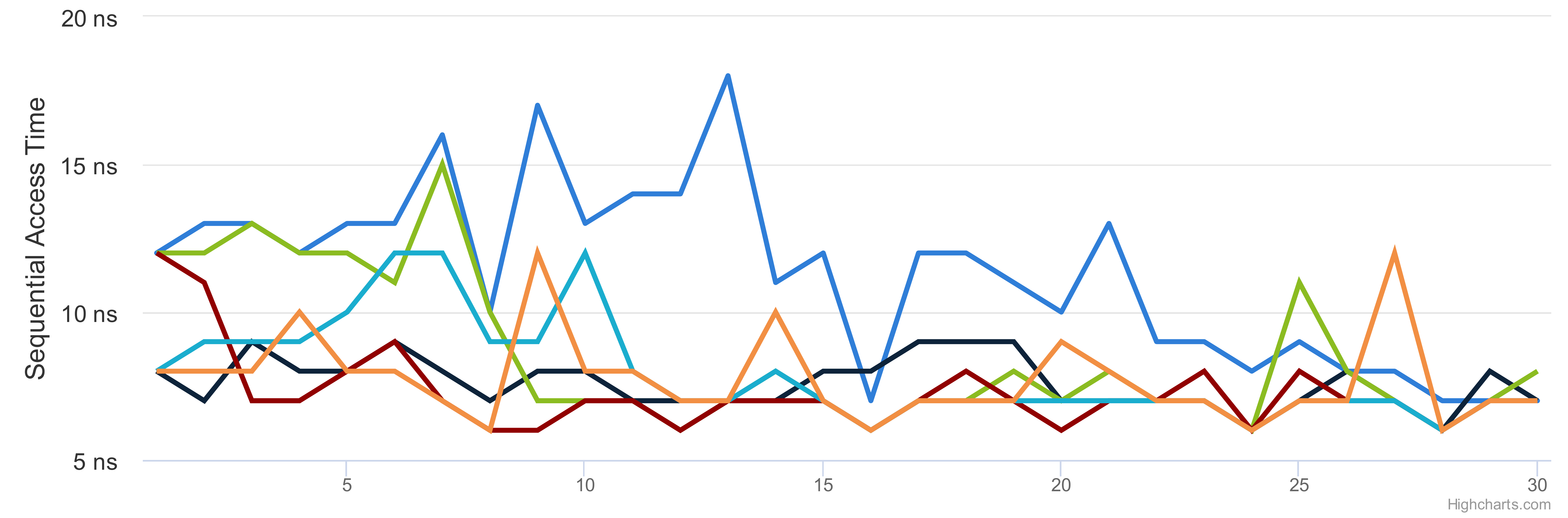}
  
  \vspace{.5cm}

  \includegraphics[width=\columnwidth,trim={0 1cm 0 0},clip]{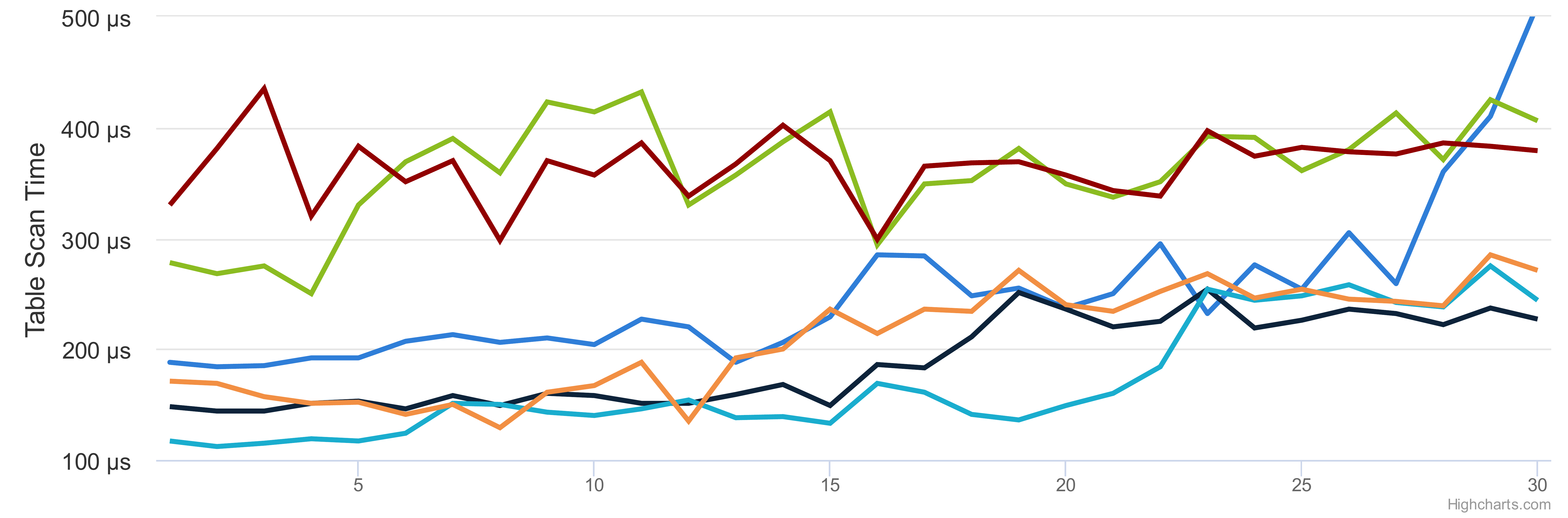}
  \caption{Performance metrics of GD Segment 1 at different deviation sizes ($x$ axis) on all six synthetic datasets.}
  \label{fig:gdv1_all_columns}
\end{figure}

%\begin{figure}[h]
%\includegraphics[width=\columnwidth]{figures/gdv1_all_columns.png}
%\caption{Compression gain (higher is better), access and table scan times (lower is better) with GD Segment 1 at different deviation sizes ($x$ axis).}
%\label{fig:gdv1_all_columns}
%\end{figure}

It is also ill-advised to assign a static default deviation size to each GD Segment variant, since their behavior changes significantly with different data distributions. \autoref{fig:gdv1_all_columns} shows the measured metrics of GD Segment 1 across all deviation sizes, encoding different columns. Again, the configuration that yields the highest compression (MC) is different than the best deviation size for late materialization. Furthermore, we cannot predict which size is best for different configuration of weights or how they compare to each other (e.g., the best deviation for maximum compression is often smaller than the one for late materialization). For GD segments, the concrete data determines performance factors, therefore running the diagnostics cannot be skipped if we want to optimize for a certain goal.

\autoref{table:gd_segment_standalone_results} and \autoref{table:gd_segment_standalone_results_2} lists the complete result set of encoding the six synthetic datasets with different segments. For GD Segments, we report the best deviation size for each of four weight distributions shown on \autoref{fig:encoding_weights}, including Early Materialization (EM) and Equal weights (EQ) in addition to MC and LM. We make the following observations.

% Replace bullet-point list with a regular paragraph to help content flow around the big tables and figures
Dictionary Encoding significantly inflates the data when all values are different, but it still achieves consistently low access speeds due to the very few memory accesses and simple algorithm for dereferencing an offset. It excels at datasets with low cardinality, achieving very high compression and speed at the same time.
Patched Frame-of-Reference is the fastest in data access in almost every case, and its size reduction is also among the best for most data distributions. An excellent choice as the default compressor in column stores.
The table scan performance of GD Segment 4 is the best across all encoders in nearly every segment, since it is heavily optimized for this operation, at the expense of access speed and compression. 
GD Segment 1 achieves the highest compression (when optimized for this metric alone) in 3 out of 6 segments. Its random and sequential access performance are the best among different GD Segment variants, since it requires the fewest memory accesses during dereferencing. Its table scan performance is very similar to Dictionary Encoding with almost all data distributions.
LZ4 achieves very high compression levels (unless presented with the notoriously uncompressible uniform random data), but it lacks efficient random access and fast table scans. This method is a reasonable default for segments that are very rarely accessed.

Note, that LZ4 and GD Segment 4 appear to have adequate sequential access speeds, but this is only a side-effect of our measurement methodology. Since these segment types do not support constant-time random or sequential access, a full decompression is performed at the beginning of the access tests and measurements read the decompressed regular vector. Therefore, the cost of the expensive one-time decompression is amortized by the 65535 extremely fast direct memory accesses. As a result, the sequential access column of GD Segment 4 and LZ4 practically only shows their decompression speeds.

\subsection{Benchmarks in Hyrise}

\begin{figure*}[th]
\subfloat[TPC-H (scale factor 5)]{%
  \includegraphics[width=\textwidth]{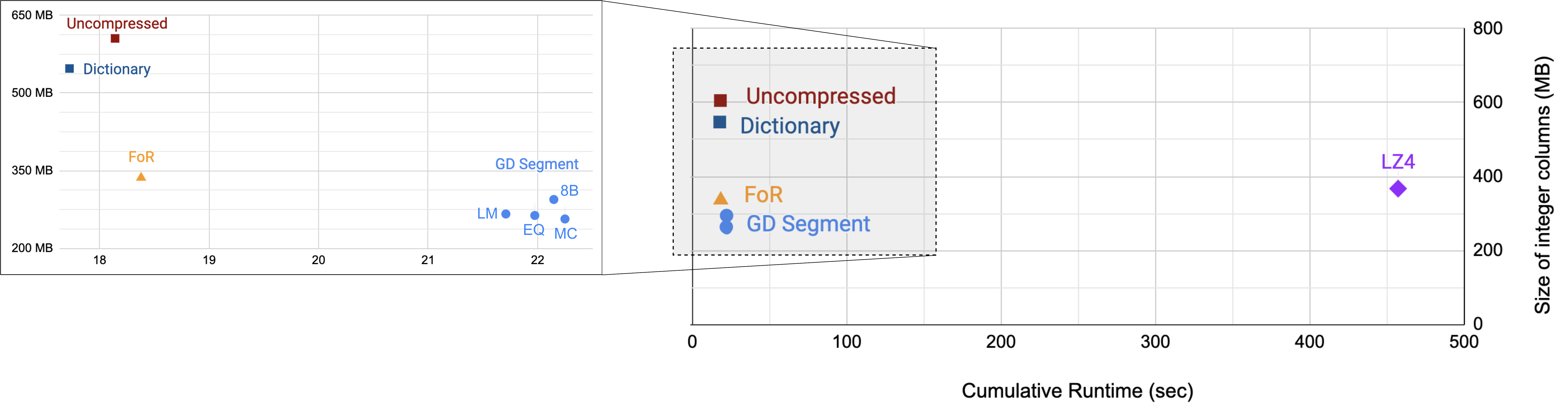}
}

\subfloat[Join Order Benchmark]{%
  \includegraphics[width=\textwidth]{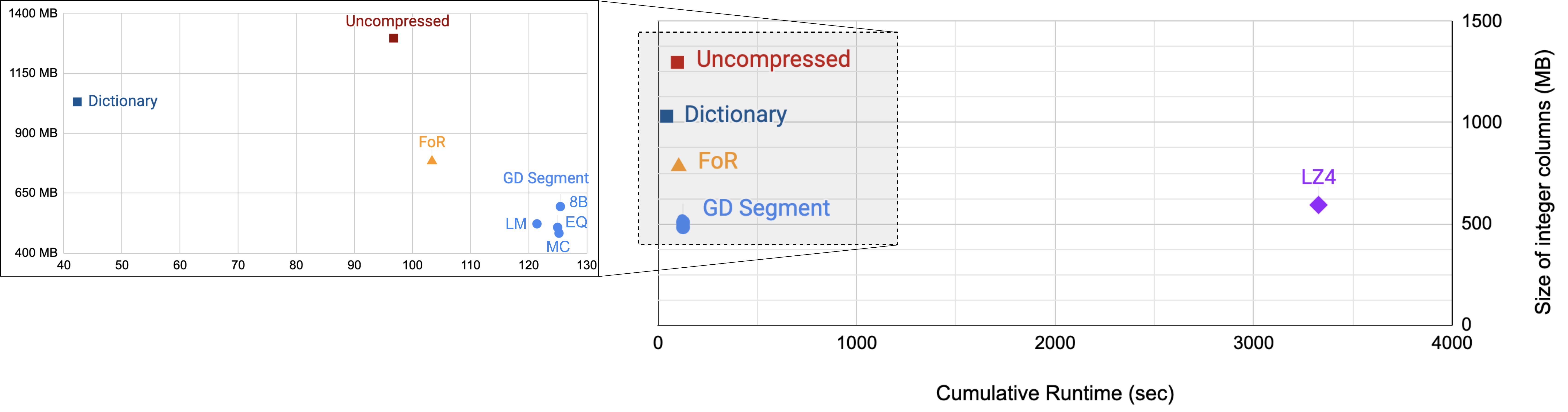}
}

\caption{Size of integer columns and total runtimes of analytical benchmarks in Hyrise, with different segment encodings.}
\label{fig:benchmark_results}
\end{figure*}

To observe the real-world performance of GD Segments and see the effects of different selection criteria for the best deviation size, we implemented GD Segment 1 as a segment encoder in Hyrise, and ran two industry-standard benchmarks: TPC-H and Join Order Benchmark (JOB). 
We chose Hyrise for this evaluation because it satisfies all our expectations about handling segments:
\begin{itemize}
    \item Segments store values of a single type.
    \item Segment encoding is a non-blocking background operation.
    \item Encoded segments are immutable.
    \item New segment encoders can be added relatively easily via an extensible framework.
\end{itemize}
 
Hyrise is a late materialization database where lists of segment offsets are passed between query nodes, and iterators are used as the interface between the query engine and segment encoders. As a result, segment performance (e.g., query performance with a given segment type) is mainly determined by the speed of iterator dereferencing. We decided to integrate GD Segment version 1, because it has the best random and sequential access performance across the four variants. When a GD segment is first encoded, we run all performance measurements described earlier and store the results in a local JSON file. Subsequent encodings of the same segment (e.g., running TPC-H again) simply load the metrics from the disk instead of having to run all tests again. The relative weights of compression and speeds are also read from a local configuration file and used by the encoder on the fly when selecting the deviation size of the segment. We measure different weight distributions for GD segments by changing the weights in the config file and re-running the benchmark. The configurations tested in the benchmarks and marked on the figures are the following: fixed 8-bit deviation size (\textit{8B}), Late Materialization (\textit{LM}), Maximum Compression (\textit{MC}) and Equal Weights (\textit{EQ}).

Note, that the built-in Frame-of-Reference encoder in Hyrise actually implements the PFoR algorithm with a block size of 2048, but it is called FoR encoder in the source code. Therefore, we will also refer to it the same way in our evaluations. Columns that cannot be encoded with the tested encoder (e.g., float and string columns with FoR and GD) are left uncoded. We report the total compressed size of integer columns and the sum of average query execution times as the cumulative benchmark runtime. All of the computation done for these benchmarks was performed on the UCloud\footnote{\href{https://docs.cloud.sdu.dk}{https://docs.cloud.sdu.dk}} interactive HPC system, which is managed by the eScience Center at the University of Southern Denmark. All measurements were performed in single-threaded mode. 

\autoref{fig:benchmark_results} shows the achieved compression and runtime of TPC-H at scale factor 5 and Join Order Benchmark. The two results show similar patterns of encoder performance. 
Dictionary encoding is the worst compressor in both benchmarks, achieving only 10\% reduction in TPC-H and 21\% in JOB, but it is the fastest as well. According to our measurements, 2\% faster than unencoded segments in TPC-H and 56\% faster in Join Order Benchmark. This makes Dictionary Segment the only contender for Hyrise that simultaneously decreases data size and makes queries faster. 

However, if the query speed with unencoded segments is satisfactory for a certain use-case and decreasing the memory footprint is more important, the other three encoders offer better compression with different levels of performance penalty. Frame-of-Reference achieves 44\% (TPC-H) and 40\% (JOB) size reductions of integer columns for only 1\% and 7\% higher query times. Traditionally the only other option was to almost completely sacrifice query performance for similar or slightly better compression, using LZ4. It reduces the memory consumption by 39\% (TPC-H) and 54\% (JOB) for a staggering 24x-33x decline in query speed. This performance makes LZ4 a viable option only on rarely queried columns. 

GD Segments provide a new trade-off between query performance and compression, which is close to the speed of FoR with better size reduction than LZ4. In TPC-H, GD segment variants achieve compression between 51\% and 58\%, the highest among all encoders and 5-6x better than Dictionary encoding. In terms of query performance, GD is 20-23\% worse than unencoded segments, depending on the relative importance of metrics that guided the deviation size selection. Results for GD are similar in the Join Order Benchmark as well, with slightly higher compression (54-63\%) and lower performance (26-30\%). 
The different weight configurations used to determine the deviation size work as intended. The \textit{Late Materialization} setting yielded the fastest 
GD Segment in both benchmarks, while \textit{Maximum Compression} is indeed the one with the lowest total memory consumption. As expected, the fixed 8-bit deviation is worse than both of them.

% from https://www.latex-tables.com/
% Replace 'μs' with '$\mu$s' after copying the table
\begin{table*}
  \centering
  \caption{Detailed results of standalone evaluation of GD Segments. Values in bold indicate the best of the column. MC, EQ, LM and EM in the \textit{Best Deviation Size} (Best Dev.) column marks which segment encoding (and which deviation size in case of GD) yields the best overall choice for the given optimization goal.}
  \label{table:gd_segment_standalone_results}
  \arrayrulecolor{black}
  \begin{tabular}{|c|c|c|c|c|c|c||c|c|c|c|c|}
  
  \multicolumn{1}{c}{} & \multicolumn{1}{c!{\color[rgb]{0.949,0.949,0.949}\vrule}}{} & \multicolumn{5}{c!{\color[rgb]{0.949,0.949,0.949}\vrule}}{\textbf{Uniform Random}} & \multicolumn{5}{c!{\color[rgb]{0.949,0.949,0.949}\vrule}}{\textbf{Sorted Equi-Distance}} \\
  \multicolumn{1}{c}{} & \textbf{Version} & \multicolumn{1}{c}{\textbf{Best Dev.}} & \multicolumn{1}{c}{\textbf{Comp.}} & \multicolumn{1}{c}{\textbf{Rand. Acc}} & \multicolumn{1}{c}{\textbf{Seq. Acc}} & \textbf{Scan} & \multicolumn{1}{c}{\textbf{\textbf{Dev.}}} & \multicolumn{1}{c}{\textbf{\textbf{Comp.}}} & \multicolumn{1}{c}{\textbf{\textbf{Rand. Acc}}} & \multicolumn{1}{c}{\textbf{\textbf{Seq. Acc}}} & \multicolumn{1}{c!{\color[rgb]{0.949,0.949,0.949}\vrule}}{\textbf{\textbf{Scan}}} \\ 
  \hline
  \rowcolor[rgb]{0.937,0.937,0.937} {\cellcolor[rgb]{0.937,0.937,0.937}} & MC & 28~(MC) & \textbf{3\%} & 14~ns & 10~ns & 483 $\mu$s & 17 & 41\% & 8~ns & 7~ns & 183~$\mu$s \\ 
  \hhline{|>{\arrayrulecolor[rgb]{0.937,0.937,0.937}}->{\arrayrulecolor{black}}------||-----|}
  \rowcolor[rgb]{0.937,0.937,0.937} {\cellcolor[rgb]{0.937,0.937,0.937}} & EQ & {\cellcolor[rgb]{0.937,0.937,0.937}} & {\cellcolor[rgb]{0.937,0.937,0.937}} & {\cellcolor[rgb]{0.937,0.937,0.937}} & {\cellcolor[rgb]{0.937,0.937,0.937}} & {\cellcolor[rgb]{0.937,0.937,0.937}} & {\cellcolor[rgb]{0.937,0.937,0.937}} & {\cellcolor[rgb]{0.937,0.937,0.937}} & {\cellcolor[rgb]{0.937,0.937,0.937}} & {\cellcolor[rgb]{0.937,0.937,0.937}} & {\cellcolor[rgb]{0.937,0.937,0.937}} \\ 
  \hhline{|>{\arrayrulecolor[rgb]{0.937,0.937,0.937}}->{\arrayrulecolor{black}}->{\arrayrulecolor[rgb]{0.937,0.937,0.937}}----->{\arrayrulecolor{black}}||>{\arrayrulecolor[rgb]{0.937,0.937,0.937}}----->{\arrayrulecolor{black}}|}
  \rowcolor[rgb]{0.937,0.937,0.937} {\cellcolor[rgb]{0.937,0.937,0.937}} & LM & {\cellcolor[rgb]{0.937,0.937,0.937}} & {\cellcolor[rgb]{0.937,0.937,0.937}} & {\cellcolor[rgb]{0.937,0.937,0.937}} & {\cellcolor[rgb]{0.937,0.937,0.937}} & {\cellcolor[rgb]{0.937,0.937,0.937}} & {\cellcolor[rgb]{0.937,0.937,0.937}} & {\cellcolor[rgb]{0.937,0.937,0.937}} & {\cellcolor[rgb]{0.937,0.937,0.937}} & {\cellcolor[rgb]{0.937,0.937,0.937}} & {\cellcolor[rgb]{0.937,0.937,0.937}} \\ 
  \hhline{|>{\arrayrulecolor[rgb]{0.937,0.937,0.937}}->{\arrayrulecolor{black}}->{\arrayrulecolor[rgb]{0.937,0.937,0.937}}----->{\arrayrulecolor{black}}||>{\arrayrulecolor[rgb]{0.937,0.937,0.937}}----->{\arrayrulecolor{black}}|}
  \rowcolor[rgb]{0.937,0.937,0.937} \multirow{-4}{*}{{\cellcolor[rgb]{0.937,0.937,0.937}}\textbf{GD S1}} & EM & \multirow{-3}{*}{{\cellcolor[rgb]{0.937,0.937,0.937}}24} & \multirow{-3}{*}{{\cellcolor[rgb]{0.937,0.937,0.937}}\textbf{3\% }} & \multirow{-3}{*}{{\cellcolor[rgb]{0.937,0.937,0.937}}13~ns} & \multirow{-3}{*}{{\cellcolor[rgb]{0.937,0.937,0.937}}11~ns} & \multirow{-3}{*}{{\cellcolor[rgb]{0.937,0.937,0.937}}359~$\mu$s} & \multirow{-3}{*}{{\cellcolor[rgb]{0.937,0.937,0.937}}16} & \multirow{-3}{*}{{\cellcolor[rgb]{0.937,0.937,0.937}}41\%} & \multirow{-3}{*}{{\cellcolor[rgb]{0.937,0.937,0.937}}6~ns} & \multirow{-3}{*}{{\cellcolor[rgb]{0.937,0.937,0.937}}6~ns} & \multirow{-3}{*}{{\cellcolor[rgb]{0.937,0.937,0.937}}182~$\mu$s} \\ 
  \hline
  \multirow{4}{*}{\textbf{GD S2}} & MC & 13 & -48\% & 31~ns & 21~ns & 353~$\mu$s & 10 & 40\% & 19~ns & 12~ns & 228~$\mu$s \\ 
  \cline{2-12}
   & EQ & \multirow{3}{*}{16} & \multirow{3}{*}{-50\%} & \multirow{3}{*}{19~ns} & \multirow{3}{*}{11~ns} & \multirow{3}{*}{369~$\mu$s} & \multirow{2}{*}{8} & \multirow{2}{*}{39\%} & \multirow{2}{*}{15~ns} & \multirow{2}{*}{10~ns} & \multirow{2}{*}{225~$\mu$s} \\ 
  \cline{2-2}
   & LM &  &  &  &  &  &  &  &  &  &  \\ 
  \cline{2-2}\cline{8-12}
   & EM &  &  &  &  &  & 1 & -50\% & 24~ns & 9~ns & 151~$\mu$s \\ 
  \hline
  \rowcolor[rgb]{0.937,0.937,0.937} {\cellcolor[rgb]{0.937,0.937,0.937}} & MC & 20 & -20\% & 35~ns & 23~ns & 346~$\mu$s & 7 & 20\% & 30~ns & 14~ns & 201~$\mu$s \\ 
  \hhline{|>{\arrayrulecolor[rgb]{0.937,0.937,0.937}}->{\arrayrulecolor{black}}------||-----|}
  \rowcolor[rgb]{0.937,0.937,0.937} {\cellcolor[rgb]{0.937,0.937,0.937}} & EQ & 23 & -25\% & 31~ns & 23~ns & 311~$\mu$s & {\cellcolor[rgb]{0.937,0.937,0.937}} & {\cellcolor[rgb]{0.937,0.937,0.937}} & {\cellcolor[rgb]{0.937,0.937,0.937}} & {\cellcolor[rgb]{0.937,0.937,0.937}} & {\cellcolor[rgb]{0.937,0.937,0.937}} \\ 
  \hhline{|>{\arrayrulecolor[rgb]{0.937,0.937,0.937}}->{\arrayrulecolor{black}}------||>{\arrayrulecolor[rgb]{0.937,0.937,0.937}}----->{\arrayrulecolor{black}}|}
  \rowcolor[rgb]{0.937,0.937,0.937} {\cellcolor[rgb]{0.937,0.937,0.937}} & LM & 24 & -28\% & 29~ns & 20~ns & 352~$\mu$s & \multirow{-2}{*}{{\cellcolor[rgb]{0.937,0.937,0.937}}18} & \multirow{-2}{*}{{\cellcolor[rgb]{0.937,0.937,0.937}}-13\%} & \multirow{-2}{*}{{\cellcolor[rgb]{0.937,0.937,0.937}}14~ns} & \multirow{-2}{*}{{\cellcolor[rgb]{0.937,0.937,0.937}}10~ns} & \multirow{-2}{*}{{\cellcolor[rgb]{0.937,0.937,0.937}}187~$\mu$s} \\ 
  \hhline{|>{\arrayrulecolor[rgb]{0.937,0.937,0.937}}->{\arrayrulecolor{black}}------||-----|}
  \rowcolor[rgb]{0.937,0.937,0.937} \multirow{-4}{*}{{\cellcolor[rgb]{0.937,0.937,0.937}}\textbf{GD S3}} & EM & 23 (EM) & -25\% & 30~ns & 23~ns & 311~$\mu$s & 13 & 6\% & 22~ns & 11~ns & 173~$\mu$s \\ 
  \hline
  \multirow{4}{*}{\textbf{GD S4}} & MC & 21 & -19\% & 50~ns & 7~ns & 89~$\mu$s & 8 & 23\% & 34~ns & 6~ns & 71~$\mu$s \\ 
  \cline{2-12}
   & EQ & \multirow{2}{*}{25} & \multirow{2}{*}{-28\%} & \multirow{2}{*}{47~ns} & \multirow{2}{*}{6~ns} & \multirow{2}{*}{76~$\mu$s} & \multirow{3}{*}{\begin{tabular}[c]{@{}c@{}}26\\(EM)\end{tabular}} & \multirow{3}{*}{-30\%} & \multirow{3}{*}{33~ns} & \multirow{3}{*}{5~ns} & \multirow{3}{*}{\textbf{68~$\mu$s}} \\ 
  \cline{2-2}
   & LM &  &  &  &  &  &  &  &  &  &  \\ 
  \cline{2-7}
   & EM & 26 & -31\% & 47~ns & 6~ns & \textbf{75~$\mu$s} &  &  &  &  &  \\ 
  \hline
  \rowcolor[rgb]{0.937,0.937,0.937} \multicolumn{2}{|c|}{\textbf{Dictionary}} &  & -50\% & 8 ns & 7 ns & 421~$\mu$s &  & -100\% & 8~ns & \textbf{4~ns} & 189~$\mu$s \\ 
  \hline
  \multicolumn{2}{|c|}{\textbf{FoR}} &  & 0\% & \textbf{7 ns} & 6 ns & 476~$\mu$s & EQ, LM & 50\% & \textbf{4~ns} & \textbf{4~ns} & 301~$\mu$s \\ 
  \hline
  \rowcolor[rgb]{0.937,0.937,0.937} \multicolumn{2}{|c|}{\textbf{LZ4}} & EQ, LM & 0\% & 10 ns & \textbf{3 ns} & 262~$\mu$s & MC & \textbf{65}\% & 112~ns & 13~ns & 779~$\mu$s \\ 
  \hline
  \multicolumn{2}{c}{} & \multicolumn{5}{c}{\begin{tabular}[c]{@{}c@{}}\textbf{\textbf{}}\\\textbf{\textbf{Years}}\end{tabular}} & \multicolumn{5}{c}{\begin{tabular}[c]{@{}c@{}}\textbf{}\\\textbf{Months}\end{tabular}} \\ 
  \arrayrulecolor[rgb]{0.949,0.949,0.949}\cline{12-12}
  \multicolumn{1}{c}{} & \textbf{\textbf{Version}} & \multicolumn{1}{c}{\textbf{\textbf{Best~}\textbf{Dev.}}} & \multicolumn{1}{c}{\textbf{\textbf{Comp.}}} & \multicolumn{1}{c}{\textbf{\textbf{Rand. Acc}}} & \multicolumn{1}{c}{\textbf{\textbf{Seq. Acc}}} & \textbf{\textbf{Scan}} & \multicolumn{1}{c}{\textbf{\textbf{\textbf{\textbf{Dev.}}}}} & \multicolumn{1}{c}{\textbf{\textbf{\textbf{\textbf{Comp.}}}}} & \multicolumn{1}{c}{\textbf{\textbf{\textbf{\textbf{Rand. Acc}}}}} & \multicolumn{1}{c}{\textbf{\textbf{\textbf{\textbf{Seq. Acc}}}}} & \multicolumn{1}{c!{\color[rgb]{0.949,0.949,0.949}\vrule}}{\textbf{\textbf{\textbf{\textbf{Scan}}}}} \\ 
  \arrayrulecolor{black}\hline
  \rowcolor[rgb]{0.937,0.937,0.937} {\cellcolor[rgb]{0.937,0.937,0.937}} & MC & 6 (MC) & \textbf{75\%} & 26 ns & 20 ns & 408 $\mu$s & 3~(MC) & \textbf{87\%} & 6 ns & 6 ns & 415 $\mu$s \\ 
  \hhline{|>{\arrayrulecolor[rgb]{0.937,0.937,0.937}}->{\arrayrulecolor{black}}------||-----|}
  \rowcolor[rgb]{0.937,0.937,0.937} {\cellcolor[rgb]{0.937,0.937,0.937}} & EQ & {\cellcolor[rgb]{0.937,0.937,0.937}} & {\cellcolor[rgb]{0.937,0.937,0.937}} & {\cellcolor[rgb]{0.937,0.937,0.937}} & {\cellcolor[rgb]{0.937,0.937,0.937}} & {\cellcolor[rgb]{0.937,0.937,0.937}} & 4 & 84\% & 6 ns & 6 ns & 292 $\mu$s \\ 
  \hhline{|>{\arrayrulecolor[rgb]{0.937,0.937,0.937}}->{\arrayrulecolor{black}}->{\arrayrulecolor[rgb]{0.937,0.937,0.937}}----->{\arrayrulecolor{black}}||-----|}
  \rowcolor[rgb]{0.937,0.937,0.937} {\cellcolor[rgb]{0.937,0.937,0.937}} & LM & {\cellcolor[rgb]{0.937,0.937,0.937}} & {\cellcolor[rgb]{0.937,0.937,0.937}} & {\cellcolor[rgb]{0.937,0.937,0.937}} & {\cellcolor[rgb]{0.937,0.937,0.937}} & {\cellcolor[rgb]{0.937,0.937,0.937}} & 8 & 72\% & 5 ns & 6 ns & 297 $\mu$s \\ 
  \hhline{|>{\arrayrulecolor[rgb]{0.937,0.937,0.937}}->{\arrayrulecolor{black}}->{\arrayrulecolor[rgb]{0.937,0.937,0.937}}----->{\arrayrulecolor{black}}||-----|}
  \rowcolor[rgb]{0.937,0.937,0.937} \multirow{-4}{*}{{\cellcolor[rgb]{0.937,0.937,0.937}}\textbf{\textbf{GD S1}}} & EM & \multirow{-3}{*}{{\cellcolor[rgb]{0.937,0.937,0.937}}16} & \multirow{-3}{*}{{\cellcolor[rgb]{0.937,0.937,0.937}}47\%} & \multirow{-3}{*}{{\cellcolor[rgb]{0.937,0.937,0.937}}5 ns} & \multirow{-3}{*}{{\cellcolor[rgb]{0.937,0.937,0.937}}6 ns} & \multirow{-3}{*}{{\cellcolor[rgb]{0.937,0.937,0.937}}314 $\mu$s} & 4 & 84\% & 6 ns & 6 ns & 292 $\mu$s \\ 
  \hline
  \multirow{4}{*}{\textbf{\textbf{GD S2}}} & MC & 5 & 75\% & 12 ns & 13 ns & 272 $\mu$s & 3 & 87\% & 6 ns & 6 ns & 328 $\mu$s \\ 
  \cline{2-12}
   & EQ & \multirow{3}{*}{16} & \multirow{3}{*}{75\%} & \multirow{3}{*}{6 ns} & \multirow{3}{*}{6 ns} & \multirow{3}{*}{257 $\mu$s} & \multirow{3}{*}{4} & \multirow{3}{*}{87\%} & \multirow{3}{*}{6 ns} & \multirow{3}{*}{6 ns} & \multirow{3}{*}{232 $\mu$s} \\ 
  \cline{2-2}
   & LM &  &  &  &  &  &  &  &  &  &  \\ 
  \cline{2-2}
   & EM &  &  &  &  &  &  &  &  &  &  \\ 
  \hline
  \rowcolor[rgb]{0.937,0.937,0.937} {\cellcolor[rgb]{0.937,0.937,0.937}} & MC & 4 & 75\% & 14 ns & 14 ns & 290 $\mu$s & {\cellcolor[rgb]{0.937,0.937,0.937}} & {\cellcolor[rgb]{0.937,0.937,0.937}} & {\cellcolor[rgb]{0.937,0.937,0.937}} & {\cellcolor[rgb]{0.937,0.937,0.937}} & {\cellcolor[rgb]{0.937,0.937,0.937}} \\ 
  \hhline{|>{\arrayrulecolor[rgb]{0.937,0.937,0.937}}->{\arrayrulecolor{black}}------||>{\arrayrulecolor[rgb]{0.937,0.937,0.937}}----->{\arrayrulecolor{black}}|}
  \rowcolor[rgb]{0.937,0.937,0.937} {\cellcolor[rgb]{0.937,0.937,0.937}} & EQ & {\cellcolor[rgb]{0.937,0.937,0.937}} & {\cellcolor[rgb]{0.937,0.937,0.937}} & {\cellcolor[rgb]{0.937,0.937,0.937}} & {\cellcolor[rgb]{0.937,0.937,0.937}} & {\cellcolor[rgb]{0.937,0.937,0.937}} & {\cellcolor[rgb]{0.937,0.937,0.937}} & {\cellcolor[rgb]{0.937,0.937,0.937}} & {\cellcolor[rgb]{0.937,0.937,0.937}} & {\cellcolor[rgb]{0.937,0.937,0.937}} & {\cellcolor[rgb]{0.937,0.937,0.937}} \\ 
  \hhline{|>{\arrayrulecolor[rgb]{0.937,0.937,0.937}}->{\arrayrulecolor{black}}->{\arrayrulecolor[rgb]{0.937,0.937,0.937}}----->{\arrayrulecolor{black}}||>{\arrayrulecolor[rgb]{0.937,0.937,0.937}}----->{\arrayrulecolor{black}}|}
  \rowcolor[rgb]{0.937,0.937,0.937} {\cellcolor[rgb]{0.937,0.937,0.937}} & LM & {\cellcolor[rgb]{0.937,0.937,0.937}} & {\cellcolor[rgb]{0.937,0.937,0.937}} & {\cellcolor[rgb]{0.937,0.937,0.937}} & {\cellcolor[rgb]{0.937,0.937,0.937}} & {\cellcolor[rgb]{0.937,0.937,0.937}} & {\cellcolor[rgb]{0.937,0.937,0.937}} & {\cellcolor[rgb]{0.937,0.937,0.937}} & {\cellcolor[rgb]{0.937,0.937,0.937}} & {\cellcolor[rgb]{0.937,0.937,0.937}} & {\cellcolor[rgb]{0.937,0.937,0.937}} \\ 
  \hhline{|>{\arrayrulecolor[rgb]{0.937,0.937,0.937}}->{\arrayrulecolor{black}}->{\arrayrulecolor[rgb]{0.937,0.937,0.937}}----->{\arrayrulecolor{black}}||>{\arrayrulecolor[rgb]{0.937,0.937,0.937}}----->{\arrayrulecolor{black}}|}
  \rowcolor[rgb]{0.937,0.937,0.937} \multirow{-4}{*}{{\cellcolor[rgb]{0.937,0.937,0.937}}\textbf{\textbf{GD S3}}} & EM & \multirow{-3}{*}{{\cellcolor[rgb]{0.937,0.937,0.937}}16} & \multirow{-3}{*}{{\cellcolor[rgb]{0.937,0.937,0.937}}75\%} & \multirow{-3}{*}{{\cellcolor[rgb]{0.937,0.937,0.937}}6 ns} & \multirow{-3}{*}{{\cellcolor[rgb]{0.937,0.937,0.937}}6 ns} & \multirow{-3}{*}{{\cellcolor[rgb]{0.937,0.937,0.937}}244 $\mu$s} & \multirow{-4}{*}{{\cellcolor[rgb]{0.937,0.937,0.937}}\begin{tabular}[c]{@{}>{\cellcolor[rgb]{0.937,0.937,0.937}}c@{}}4\\(EQ, EM)\end{tabular}} & \multirow{-4}{*}{{\cellcolor[rgb]{0.937,0.937,0.937}}87\%} & \multirow{-4}{*}{{\cellcolor[rgb]{0.937,0.937,0.937}}5 ns} & \multirow{-4}{*}{{\cellcolor[rgb]{0.937,0.937,0.937}}6 ns} & \multirow{-4}{*}{{\cellcolor[rgb]{0.937,0.937,0.937}}\textbf{231 $\mu$s}} \\ 
  \hline
  \multirow{4}{*}{\textbf{\textbf{GD S4}}} & MC & 1 & 46\% & 32 ns & 6 ns & 185 $\mu$s & \multirow{2}{*}{1} & \multirow{2}{*}{47\%} & \multirow{2}{*}{39 ns} & \multirow{2}{*}{5 ns} & \multirow{2}{*}{295 $\mu$s} \\ 
  \cline{2-7}
   & EQ & 2 & 44\% & 31 ns & 5 ns & \textbf{182 $\mu$s} &  &  &  &  &  \\ 
  \cline{2-12}
   & LM & 8 & 25\% & 29 ns & 5 ns & 319 $\mu$s & 16 & 0\% & 29 ns & 5 ns & 343 $\mu$s \\ 
  \cline{2-12}
   & EM & 2 & 44\% & 31 ns & 5 ns & \textbf{182 $\mu$s} & 1 & 47\% & 39 ns & 5 ns & 295 $\mu$s \\ 
  \hline
  \rowcolor[rgb]{0.937,0.937,0.937} \multicolumn{2}{|c|}{\textbf{\textbf{Dictionary}}} & EQ, LM, EM & 75\% & \textbf{4 ns} & \textbf{4 ns} & 319 $\mu$s & LM & 75\% & \textbf{4 ns} & \textbf{4 ns} & 345 $\mu$s \\ 
  \hline
  \multicolumn{2}{|c|}{\textbf{\textbf{FoR}}} & \multicolumn{1}{l|}{} & 75\% & \textbf{4 ns} & 5 ns & 354 $\mu$s &  & 75\% & 10 ns & 4 ns & 368 $\mu$s \\ 
  \hline
  \rowcolor[rgb]{0.937,0.937,0.937} \multicolumn{2}{|c|}{\textbf{\textbf{LZ4}}~} & \multicolumn{1}{l|}{} & 65\% & 97 ns & 12 ns & 711 $\mu$s &  & 84\% & 66 ns & 8 ns & 535 $\mu$s \\ 
  \hline

  \end{tabular}
\end{table*}
  
  % Second half of the standalone results table
  \begin{table*}
  \centering
  \caption{Detailed results of standalone evaluation of GD Segments on Time Series and Primary Key synthetic datasets (continuation of \ref{table:gd_segment_standalone_results})}
  \label{table:gd_segment_standalone_results_2}
  \arrayrulecolor{black}
  \begin{tabular}{|c|c|c|c|c|c|c||c|c|c|c|c|}
  
      \multicolumn{1}{c}{} & \multicolumn{1}{c}{} & \multicolumn{5}{c}{\begin{tabular}[c]{@{}c@{}}\\\textbf{Time Series}\end{tabular}} & \multicolumn{5}{c}{\begin{tabular}[c]{@{}c@{}}\textbf{}\\\textbf{Primary Key}\end{tabular}} \\
      \multicolumn{1}{c}{} & \textbf{\textbf{\textbf{\textbf{Version}}}} & \multicolumn{1}{c}{\textbf{\textbf{Best~}\textbf{\textbf{\textbf{Dev.}}}}} & \multicolumn{1}{c}{\textbf{\textbf{\textbf{\textbf{Comp.}}}}} & \multicolumn{1}{c}{\textbf{\textbf{\textbf{\textbf{Rand. Acc}}}}} & \multicolumn{1}{c}{\textbf{\textbf{\textbf{\textbf{Seq. Acc}}}}} & \textbf{\textbf{\textbf{\textbf{Scan}}}} & \multicolumn{1}{c}{\textbf{\textbf{\textbf{\textbf{\textbf{\textbf{\textbf{\textbf{Dev.}}}}}}}}} & \multicolumn{1}{c}{\textbf{\textbf{\textbf{\textbf{\textbf{\textbf{\textbf{\textbf{Comp.}}}}}}}}} & \multicolumn{1}{c}{\textbf{\textbf{\textbf{\textbf{\textbf{\textbf{\textbf{\textbf{Rand. Acc}}}}}}}}} & \multicolumn{1}{c}{\textbf{\textbf{\textbf{\textbf{\textbf{\textbf{\textbf{\textbf{Seq. Acc}}}}}}}}} & \multicolumn{1}{c!{\color[rgb]{0.949,0.949,0.949}\vrule}}{\textbf{\textbf{\textbf{\textbf{\textbf{\textbf{\textbf{\textbf{Scan}}}}}}}}} \\ 
      \hline
      \rowcolor[rgb]{0.937,0.937,0.937} {\cellcolor[rgb]{0.937,0.937,0.937}} & MC & 21 & 28\% & 8 ns & 7 ns & 168 $\mu$s & 15 & 50\% & 10 ns & 9 ns & 231 $\mu$s \\ 
      \hhline{|>{\arrayrulecolor[rgb]{0.937,0.937,0.937}}->{\arrayrulecolor{black}}------||-----|}
      \rowcolor[rgb]{0.937,0.937,0.937} {\cellcolor[rgb]{0.937,0.937,0.937}} & EQ & {\cellcolor[rgb]{0.937,0.937,0.937}} & {\cellcolor[rgb]{0.937,0.937,0.937}} & {\cellcolor[rgb]{0.937,0.937,0.937}} & {\cellcolor[rgb]{0.937,0.937,0.937}} & {\cellcolor[rgb]{0.937,0.937,0.937}} & 8 & 50\% & 9 ns & 7 ns & 139 $\mu$s \\ 
      \hhline{|>{\arrayrulecolor[rgb]{0.937,0.937,0.937}}->{\arrayrulecolor{black}}->{\arrayrulecolor[rgb]{0.937,0.937,0.937}}----->{\arrayrulecolor{black}}||-----|}
      \rowcolor[rgb]{0.937,0.937,0.937} {\cellcolor[rgb]{0.937,0.937,0.937}} & LM & {\cellcolor[rgb]{0.937,0.937,0.937}} & {\cellcolor[rgb]{0.937,0.937,0.937}} & {\cellcolor[rgb]{0.937,0.937,0.937}} & {\cellcolor[rgb]{0.937,0.937,0.937}} & {\cellcolor[rgb]{0.937,0.937,0.937}} & 16 & 47\% & 6 ns & 6 ns & 238 $\mu$s \\ 
      \hhline{|>{\arrayrulecolor[rgb]{0.937,0.937,0.937}}->{\arrayrulecolor{black}}->{\arrayrulecolor[rgb]{0.937,0.937,0.937}}----->{\arrayrulecolor{black}}||-----|}
      \rowcolor[rgb]{0.937,0.937,0.937} \multirow{-4}{*}{{\cellcolor[rgb]{0.937,0.937,0.937}}\textbf{\textbf{GD S1}}} & EM & \multirow{-3}{*}{{\cellcolor[rgb]{0.937,0.937,0.937}}16} & \multirow{-3}{*}{{\cellcolor[rgb]{0.937,0.937,0.937}}28\%} & \multirow{-3}{*}{{\cellcolor[rgb]{0.937,0.937,0.937}}7 ns} & \multirow{-3}{*}{{\cellcolor[rgb]{0.937,0.937,0.937}}6 ns} & \multirow{-3}{*}{{\cellcolor[rgb]{0.937,0.937,0.937}}167 $\mu$s} & 8 & 50\% & 9 ns & 7 ns & 139 $\mu$s \\ 
      \hline
      \multirow{4}{*}{\textbf{\textbf{GD S2}}} & MC & 12 & 26\% & 18 ns & 14 ns & 200 $\mu$s & 9 & 50\% & 13 ns & 8 ns & 145 $\mu$s \\ 
      \cline{2-12}
       & EQ & \multirow{2}{*}{16} & \multirow{2}{*}{6\%} & \multirow{2}{*}{11 ns} & \multirow{2}{*}{7 ns} & \multirow{2}{*}{189 $\mu$s} & \multirow{2}{*}{8} & \multirow{2}{*}{50\%} & \multirow{2}{*}{10 ns} & \multirow{2}{*}{7 ns} & \multirow{2}{*}{159 $\mu$s} \\ 
      \cline{2-2}
       & LM &  &  &  &  &  &  &  &  &  &  \\ 
      \cline{2-12}
       & EM & 3 & -47\% & 21 ns & 9 ns & 129 $\mu$s & 3 & 39\% & 14 ns & 8 ns & 131 $\mu$s \\ 
      \hline
      \rowcolor[rgb]{0.937,0.937,0.937} {\cellcolor[rgb]{0.937,0.937,0.937}} & MC & 12 & 1\% & 24 ns & 12 ns & 185 $\mu$s & 4 & 32\% & 24 ns & 11 ns & 201 $\mu$s \\ 
      \hhline{|>{\arrayrulecolor[rgb]{0.937,0.937,0.937}}->{\arrayrulecolor{black}}------||-----|}
      \rowcolor[rgb]{0.937,0.937,0.937} {\cellcolor[rgb]{0.937,0.937,0.937}} & EQ & {\cellcolor[rgb]{0.937,0.937,0.937}} & {\cellcolor[rgb]{0.937,0.937,0.937}} & {\cellcolor[rgb]{0.937,0.937,0.937}} & {\cellcolor[rgb]{0.937,0.937,0.937}} & {\cellcolor[rgb]{0.937,0.937,0.937}} & {\cellcolor[rgb]{0.937,0.937,0.937}} & {\cellcolor[rgb]{0.937,0.937,0.937}} & {\cellcolor[rgb]{0.937,0.937,0.937}} & {\cellcolor[rgb]{0.937,0.937,0.937}} & {\cellcolor[rgb]{0.937,0.937,0.937}} \\ 
      \hhline{|>{\arrayrulecolor[rgb]{0.937,0.937,0.937}}->{\arrayrulecolor{black}}->{\arrayrulecolor[rgb]{0.937,0.937,0.937}}----->{\arrayrulecolor{black}}||>{\arrayrulecolor[rgb]{0.937,0.937,0.937}}----->{\arrayrulecolor{black}}|}
      \rowcolor[rgb]{0.937,0.937,0.937} {\cellcolor[rgb]{0.937,0.937,0.937}} & LM & {\cellcolor[rgb]{0.937,0.937,0.937}} & {\cellcolor[rgb]{0.937,0.937,0.937}} & {\cellcolor[rgb]{0.937,0.937,0.937}} & {\cellcolor[rgb]{0.937,0.937,0.937}} & {\cellcolor[rgb]{0.937,0.937,0.937}} & {\cellcolor[rgb]{0.937,0.937,0.937}} & {\cellcolor[rgb]{0.937,0.937,0.937}} & {\cellcolor[rgb]{0.937,0.937,0.937}} & {\cellcolor[rgb]{0.937,0.937,0.937}} & {\cellcolor[rgb]{0.937,0.937,0.937}} \\ 
      \hhline{|>{\arrayrulecolor[rgb]{0.937,0.937,0.937}}->{\arrayrulecolor{black}}->{\arrayrulecolor[rgb]{0.937,0.937,0.937}}----->{\arrayrulecolor{black}}||>{\arrayrulecolor[rgb]{0.937,0.937,0.937}}----->{\arrayrulecolor{black}}|}
      \rowcolor[rgb]{0.937,0.937,0.937} \multirow{-4}{*}{{\cellcolor[rgb]{0.937,0.937,0.937}}\textbf{\textbf{GD S3}}} & EM & \multirow{-3}{*}{{\cellcolor[rgb]{0.937,0.937,0.937}}23} & \multirow{-3}{*}{{\cellcolor[rgb]{0.937,0.937,0.937}}-20\%} & \multirow{-3}{*}{{\cellcolor[rgb]{0.937,0.937,0.937}}13 ns} & \multirow{-3}{*}{{\cellcolor[rgb]{0.937,0.937,0.937}}8 ns} & \multirow{-3}{*}{{\cellcolor[rgb]{0.937,0.937,0.937}}146 $\mu$s} & \multirow{-3}{*}{{\cellcolor[rgb]{0.937,0.937,0.937}}18} & \multirow{-3}{*}{{\cellcolor[rgb]{0.937,0.937,0.937}}-6\%} & \multirow{-3}{*}{{\cellcolor[rgb]{0.937,0.937,0.937}}11 ns} & \multirow{-3}{*}{{\cellcolor[rgb]{0.937,0.937,0.937}}7 ns} & \multirow{-3}{*}{{\cellcolor[rgb]{0.937,0.937,0.937}}171 $\mu$s} \\ 
      \hline
      \multirow{4}{*}{\textbf{\textbf{GD S4}}} & MC & 12 & 11\% & 46 ns & 6 ns & 73 $\mu$s & 6 & 33\% & 41 ns & 7 ns & \textbf{79 $\mu$s} \\ 
      \cline{2-12}
       & EQ & \multirow{3}{*}{\begin{tabular}[c]{@{}c@{}}16\\(EQ,LM,EM)\end{tabular}} & \multirow{3}{*}{3\%} & \multirow{3}{*}{32 ns} & \multirow{3}{*}{\textbf{4 ns}} & \multirow{3}{*}{\textbf{69 $\mu$s}} & 8~(EM) & 28\% & 35 ns & 5 ns & 84 $\mu$s \\ 
      \cline{2-2}\cline{8-12}
       & LM &  &  &  &  &  & 16 & 0\% & 28 ns & 5 ns & 103 $\mu$s \\ 
      \cline{2-2}\cline{8-12}
       & EM &  &  &  &  &  & 8~(EM) & 28\% & 35 ns & 5 ns & 84 $\mu$s \\ 
      \hline
      \rowcolor[rgb]{0.937,0.937,0.937} \multicolumn{2}{|c|}{\textbf{\textbf{\textbf{\textbf{Dictionary}}}}} & \multicolumn{1}{l|}{} & -47\% & 6 ns & 4 ns & 145 $\mu$s &  & -100\% & 7 ns & 4 ns & 202 $\mu$s \\ 
      \hline
      \multicolumn{2}{|c|}{\textbf{\textbf{\textbf{\textbf{FoR}}}}} & \multicolumn{1}{l|}{} & 0\% & \textbf{5 ns} & 4 ns & 310 $\mu$s & LM & 50\% & \textbf{4 ns} & \textbf{4 ns} & 275 $\mu$s \\ 
      \hline
      \rowcolor[rgb]{0.937,0.937,0.937} \multicolumn{2}{|c|}{\textbf{\textbf{\textbf{\textbf{LZ4}}}}~} & MC & \textbf{57\%} & 180 ns & 20 ns & 103 $\mu$s & MC & \textbf{65\%} & 110 ns & 13 ns & 744 $\mu$s \\
      \hline
  
  \end{tabular}
  \end{table*}

\section{Conclusions and Future Work} \label{sec:conclusions}

In this paper we have introduced a new column compression family based on generalized deduplication for integer sequences, and four practical designs for segment encoders in columnar databases. They aim to optimize for both memory footprint reduction and efficient query execution, without having to decompress the whole segment. 
We have shown that the performance of the proposed segments is comparable to current state-of-the-art encoders, offering a new trade-off point between compression and query speed. Additionally, we proposed an adaptive segment encoder selection scheme for autonomous databases, based on the same diagnostic and evaluation mechanism we use to automatically select the best configuration for our segment.

Our future plans include enhancing the storage layer of Hyrise with i) the ability to collect detailed, segment-level usage statistics, including the frequency and predicate of table scans, ii) a standardized set of performance measurements to evaluate segment encoders at their full parameter space, and iii) an encoding selection framework that combines the experienced usage patterns of segments with the performance metrics of potential encoders to find the best possible one, and re-evaluates this decision when the usage patterns sufficiently change.

\section{Acknowledgement}
This work was partially financed by the SCALE-IoT project (Grant No. DFF - 7026-00042B) granted by the Danish Council for Independent Research, the Aarhus Universitets Forskningsfond Starting Grant Project AUFF-2017-FLS-7-1, and Aarhus University's DIGIT Centre. 

%\clearpage

\bibliographystyle{ACM-Reference-Format}
\bibliography{refs} 

\end{document}